\documentclass[sigconf,screen]{acmart}
\usepackage[utf8]{inputenc}
\usepackage[T1]{fontenc}
\usepackage{microtype}
\usepackage{graphicx}
\usepackage{multirow}
\usepackage{booktabs}
\usepackage{array}
\usepackage{tabularx}
\usepackage{xcolor}
\usepackage{enumitem}
\usepackage{subcaption}
\usepackage{balance}
\usepackage{listings}
\usepackage{pgfplots}
\pgfplotsset{compat=1.18}
\usepackage{pgf-pie}
\usepackage{xspace}
\usepackage{amsmath}
\usepackage{seqsplit}
\usepackage{tcolorbox}
\tcbuselibrary{breakable, skins}
\usepackage{url}
\usepackage{placeins}

\newcommand{\tool}{\textsc{PAIChecker}\xspace}

\newcommand{\eg}{\textit{e.g.}\xspace}

\newcommand{\rev}[1]{#1}
\newcommand{\finding}[1]{\vspace{0.3em}
\noindent
\fbox{\parbox{0.96\columnwidth}{\textbf{Finding:} #1}}
\vspace{0.3em}}

\setcopyright{cc}
\setcctype{by}
\acmDOI{10.1145/3832783.3837557}
\acmYear{2026}
\copyrightyear{2026}
\acmISBN{979-8-4007-2882-2/2026/10}
\acmConference[ASE '26]{Proceedings of the 41st IEEE/ACM International Conference on Automated Software Engineering}{October 12--16, 2026}{Munich, Germany}
\acmBooktitle{Proceedings of the 41st IEEE/ACM International Conference on Automated Software Engineering (ASE '26), October 12--16, 2026, Munich, Germany}
\acmSubmissionID{ase26main-p3982-p}
\received{2026-03-26}
\received[accepted]{2026-06-18}

\begin{document}
    \title{PAIChecker: Uncovering and Checking PR-Issue Misalignment in SWE-Bench-Like Benchmarks}

    \author{Manyi Wang}
    \orcid{0009-0003-0802-1674}
    \affiliation{%
      \institution{The Chinese University of Hong Kong, Shenzhen}
      \city{Shenzhen}
      \country{China}
    }
    \email{manyiwang@link.cuhk.edu.cn}

    \author{Junjielong Xu}
    \authornote{Junjielong Xu and Pinjia He are co-corresponding authors.}
    \orcid{0009-0001-1516-103X}
    \affiliation{%
      \institution{The Chinese University of Hong Kong, Shenzhen}
      \city{Shenzhen}
      \country{China}
    }
    \email{junjielongxu@link.cuhk.edu.cn}

    \author{Pinjia He}
    \authornotemark[1]
    \orcid{0000-0003-3377-8129}
    \affiliation{%
      \institution{The Chinese University of Hong Kong, Shenzhen}
      \city{Shenzhen}
      \country{China}
    }
    \email{hepinjia@cuhk.edu.cn}

    \begin{abstract}

        SWE-bench-like benchmarks are widely used for
  evaluating LLM's issue resolution capability. They typically follow a common construction pipeline: each PR (Pull Request) is paired with its linked issue by extracting issue references from the PR description; the issue description is used as the problem statement, and the PR patch serves as the test oracle. However, due to the
  inherent complexity of developing and maintaining large
  repositories, such PR-Issue pairings are often misaligned in practice.
  In this work, we systematically study SWE-bench Verified instances, finding    
  that 13.6\% exhibit misalignment across five patterns in eleven fine-grained scenarios. To enable reliable and scalable construction of those benchmarks in the future, we propose \tool, a multi-agent system for checking PR-Issue misalignment in SWE-bench-like benchmarks. Specifically, \tool adopts a three‑phase design that combines specific pattern identification, cross-agent label synthesis, and code-level validation, thereby enabling more accurate, generalizable, and progressively verified detection.                               
  Experiments on SWE-Gym \rev{and SWE-bench Multilingual} show that \tool achieves the best 
  performance across all four LLM backbones, reaching up to \rev{92.12\% and 91.67\% binary accuracy, respectively.}
    
    \end{abstract}



    \maketitle

    \section{Introduction}
    \label{sec:intro}

      The emergence of large language models (LLMs) capable of code generation and
      program repair has spurred the need for rigorous, real-world benchmarks~\cite{swebench,
      jain2024livecodebenchholisticcontaminationfree, he2025sweperf,xu_aligning_2025, deng2025swepro}. Among them, SWE-bench~\cite{swebench} has become one of the most influential benchmarks for evaluating LLMs on automated issue solving tasks~\cite{team_trae_2025,xia_agentless_2024,minisweagent,zhang_autocoderover_2024, yang_kimi-dev_2025}. SWE-bench constructs task instances from real-world
      GitHub repositories, using issue descriptions as problem statements and the corresponding pull request (PR) patches as test oracles. Following its success, several SWE-bench-like
      benchmarks have emerged to address various limitations, a non-exhaustive list includes SWE-bench
      Verified~\cite{swebenchverified}, SWE-Gym~\cite{swegym}, SWE-PolyBench~\cite{swepoly}, SWE-Smith~\cite{swesmith}, and SWE-Bench-Live~\cite{swebenchlive}. These benchmarks largely adopt the same construction
      pipeline: they filter PRs by quality criteria, then use
      regular expressions to extract linked issues from PR descriptions,
      forming PR-Issue pairs as task instances. 

    However, this construction pipeline rests on a critical
    assumption: that each PR is \textit{well-aligned} with its linked issue, i.e.,
    the PR exclusively and completely addresses the stated problem,
    and the issue fully specifies what the PR solves. In
    practice, this assumption is frequently violated. A PR may bundle fixes for
    multiple issues, serve as a follow-up to a previous incomplete
    fix, introduce new defects alongside the intended fix, or implement details clarified only in later discussions rather than the original description.
    Such misalignment renders the benchmark instance problematic:
    \textit{the problem statement may not fully describe the expected solution, or the test
    oracle may evaluate aspects unrelated to the stated problem}, directly
    undermining evaluation fairness and reliability. \rev{Beyond evaluation, the same construction pipeline also underlies training-oriented datasets for code LLMs and software
  agents. Misaligned PR-Issue pairs therefore become noisy supervision, encouraging models to learn incomplete, unrelated, or
  even defective patches rather than fixes faithful to the stated issue.}

In this paper, we present a systematic study of \textit{PR-Issue misalignment} in SWE-bench-like benchmarks, where \textit{PR-Issue misalignment} refers to discrepancies between a \textit{pull request code implementation} and the \textit{issue description} it is linked to in the dataset.
We manually analyze all 500 instances in SWE-bench Verified~\cite{swebenchverified} and find that 13.6\% exhibit misalignment. We further classify these misalignments into five patterns spanning 11 fine-grained scenarios. To quantify the impact of misalignment, we correlate it with agent resolution rates and find that 41.2\% of instances never resolved by any of the 131 leaderboard agents are misaligned.

These findings motivate the need for automated detection and
categorization of PR–Issue misalignment at scale. However,
detecting misalignment directly from the code implementation and
issue description is challenging, as the intent reflected in code is
inherently ambiguous. This ambiguity arises from the nature of
code changes: in complex cases, the intended behavior may depend on modifications spanning hundreds or thousands of lines,
while in simpler cases, the implemented behavior may still be incomplete or defective. By contrast, PR-side textual artifacts, such as PR descriptions and discussions, often summarize changes at a level of abstraction similar to issues~\cite{prtemp1,prtemp2,prtemp3}, and their shared natural-language form makes the comparison between them far more straightforward and reliable. This asymmetry between textual artifacts and code motivates the key design principle of our approach, namely the \textit{text-driven, code-validation} principle, which first assesses alignment based on textual evidence and then validates it against code.
 
According to this principle, the task is well suited to LLM reasoning and tool-augmented agents, because it is fundamentally a semantic comparison task. We therefore adapt Mini-SWE-Agent~\cite{minisweagent}, augmenting it with task-specific prompts and GitHub API access, as a representative general-purpose agent baseline. However, this approach has three limitations:~(1) it lacks specialization: different misalignment patterns require different inputs and reasoning workflows, but a single prompt conflates these heterogeneous requirements and often misses subtle yet critical evidence. ~(2) it has limited generalizability: when no predefined pattern applies, it tends to force-fit the instance into the nearest known category, even though real-world misalignments often fall outside the taxonomy. ~(3) it lacks verification mechanism: its predicted label may not match its own textual reasoning as well as the code implementation. We illustrate these limitations in detail in Section~\ref{sec:motivation}.

To address these limitations, we propose \tool
(\textbf{P}R-\textbf{I}ssue \textbf{A}lignment Checker), a
three-phase multi-agent framework guided by a
\textit{text-driven, code-validation} principle. The first two
phases focus on textual analysis, and the third performs
code-level verification. Specifically, Phase I addresses
limitation(1) through three specialized subagents, each with a
focused artifact subset and tailored workflow. Phase II addresses limitation(2) via a coordinator that synthesizes cross‑subagent evidence to identify misalignment beyond known patterns. For limitation(3), \tool introduces a two-level
\textit{self-correction} mechanism: Phase~II re-verifies each
label against its textual evidence, and Phase~III validates it
against the actual code implementation.                                               

    We evaluate \tool on SWE-Gym~\cite{swegym} \rev{and SWE-bench Multilingual~\cite{swebench}},
      comparing it against \rev{three prompting baselines, and four agent-framework baselines.} Using
      four state-of-the-art LLMs as backbones (GPT-5.3 Codex~\cite{gpt53codex}, Qwen-3.5 Plus~\cite{qwen35},
      Gemini-3.1-Pro Preview~\cite{gemini31pro}, and Claude-Sonnet-4.6~\cite{claudesonnet46}), \tool achieves the best 
    performance across all four LLM backbones, reaching up to 92.12\% binary     
    accuracy and 84.66\% exact match, outperforming the strongest baseline by      
    \rev{5.13--12.39 accuracy points and 9.02--17.76 exact-match points on SWE-Gym, and by 2.33--5.67 accuracy points and 3.66--8.00 exact-match points on SWE-bench Multilingual}. Ablation studies confirm each component's contribu\nolinebreak tion.

    In summary, this paper makes the following contributions:
\begin{itemize}[leftmargin=*]
        \item \textbf{Problem Formulation.} We identify PR-Issue                          
            misalignment as a systematic construction problem in                       
            SWE-bench-like \nolinebreak bench\nolinebreak marks.
                                                                                           
      \item \textbf{Empirical Study.} We analyze all 500 SWE-bench                       
          Verified instances, deriving a taxonomy of five patterns with                    
          11 scenarios and finding 13.6\% misaligned                                       
          (Section~\ref{sec:preliminary}).                                                 
   
      \item \textbf{Detection Framework.} We propose \tool, a                              
          multi-agent framework for automated detection and                              
          categorization of PR-Issue misalignment
          (Section~\ref{sec:method}).
                                                                                           
        \item \textbf{Evaluation.} Experiments on SWE-Gym \rev{and SWE-bench Multilingual} with four
            state-of-the-art LLMs demonstrate the effectiveness of \tool.
          (Sec\nolinebreak tions~\ref{sec:results}).
                                                                                           
      \item \textbf{Data Availability.} We release all annotated data                     
          and \tool to support future benchmark curation
          ~\cite{replication}.                     
  \end{itemize}

    \section{Preliminary Study and Motivation}
    \label{sec:preliminary} In this section, we systematically investigate the
    existence, prevalence, and impact of PR-Issue misalignment in SWE-bench-like
    benchmarks. Our findings provide the empirical foundation and motivation for
    our automated misalignment checker \tool. This study focuses on two research
    questions:
    \begin{itemize}[leftmargin=*]
        \item \textbf{RQ1 (Taxonomy and Prevalence):} What are the types of PR-Issue
            misalignments, and how prevalent are they?

        \item \textbf{RQ2 (Impact):} How does PR-Issue misalignment affect agent
            evaluation reliability?
    \end{itemize}

    \subsection{Study Design}
    \label{sec:study_design}

    \subsubsection{Dataset}
    \label{sec:study_dataset} We conduct our study on SWE-bench Verified~\cite{swebenchverified},
  a human-validated subset of 500 SWE-Bench~\cite{swebench} instances.                     
  We select it for two reasons: (1)~its human validation represents a                      
  high-quality baseline, so misalignments found here likely reflect                        
  broader issues in other SWE-bench-like benchmarks; (2)~its official                      
  experiments repository~\cite{sweexperiments} provides per-instance                       
  resolution data for 131 leaderboard agents, enabling quantitative                        
  analysis for RQ2.

    \begin{table}[t]
        \centering
        \caption{Taxonomy of PR-Issue misalignment patterns in SWE-bench Verified (68/500 instances). Instances may exhibit multiple patterns.}
        \label{tab:misalignment_patterns}
        \footnotesize
        \begin{tabularx}{\columnwidth}{@{}>{\raggedright\arraybackslash}p{0.3\columnwidth}X@{}}
            \toprule
            \textbf{Pattern} & \textbf{Scenario} \\
            \midrule
            \multirow{4}{=}{\textbf{SC:} PR Scope Creep}
                & \textbf{SC-1\,(12):} Resolves multiple issues; only one specified \\
                & \textbf{SC-2\,(2):} Adds features beyond issue description \\
                & \textbf{SC-3\,(2):} Bundles fixes for bugs not in the issue \\
                & \textbf{SC-4\,(6):} Extra patches for other issues alongside fix \\
            \midrule
            \multirow{2}{=}{\textbf{DP:} Defective PR}
                & \textbf{DP-1\,(14):} Introduces new bugs requiring follow-up fixes \\
                & \textbf{DP-2\,(16):} Incomplete solution requiring follow-up \\
            \midrule
            \multirow{2}{=}{\textbf{IS:} Incomplete Specification}
                & \textbf{IS-1\,(12):} Details supplemented by reporter in discussion \\
                & \textbf{IS-2\,(6):} Addresses new problem from later discussion \\
            \midrule
            \multirow{2}{=}{\textbf{FP:} Follow-up PR}
                & \textbf{FP-1\,(2):} Fixes bugs introduced by a previous PR \\
                & \textbf{FP-2\,(1):} Supplements a prior PR \\
            \midrule
            \textbf{UL:} Unspecified Literal
                & \textbf{UL-1\,(1):} Asserts exact literals (\eg messages) not in issue \\
            \bottomrule
        \end{tabularx}
    \end{table}

    \subsubsection{Method}
    \label{sec:study_method}

    \newcommand{\artifactlabel}[1]{%
      \noindent\textbf{\small #1}\vspace{2pt}\hrule\vspace{4pt}}
    \begin{figure*}[t]
    \vspace{2mm}
    \captionsetup{justification=raggedright}
      \centering
      \begin{subfigure}[t]{0.235\textwidth}
        \includegraphics[pagebox=cropbox,clip,width=\linewidth]{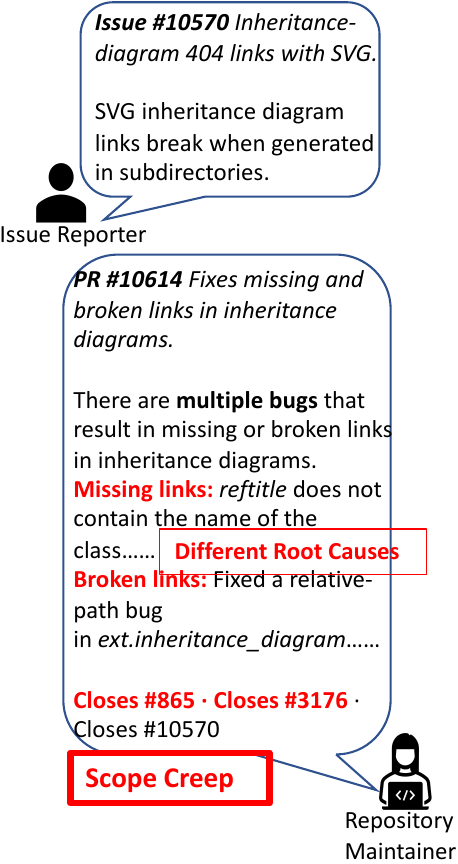}
        \caption{SC-1(sphinx\-doc\_\_sphinx-\allowbreak 10614): The PR closed three distinct issues; the problem statement covers only one. }
        \label{fig:sc1_example}
      \end{subfigure}
      \hfill
      \begin{subfigure}[t]{0.235\textwidth}
        \includegraphics[pagebox=cropbox,clip,width=\linewidth]{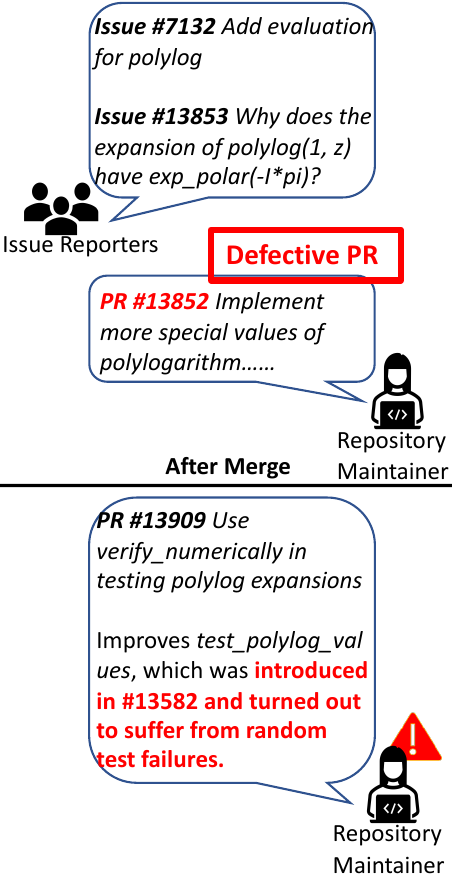}
        \caption{DP-1(sympy\_\_sympy-13852): The PR introduces random test failures that require a follow‑up fix.}
        \label{fig:dp_example}
      \end{subfigure}
      \hfill
      \begin{subfigure}[t]{0.235\textwidth}
        \includegraphics[pagebox=cropbox,clip,width=\linewidth]{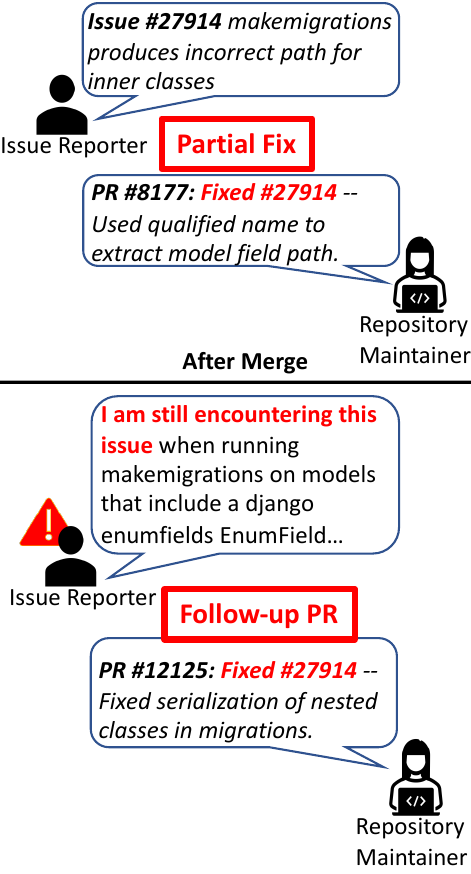}
        \caption{FP-2(django\_\_django-12125): The PR addresses only the remaining bug not the full issue.}
        \label{fig:fp_example}
      \end{subfigure}
      \hfill
      \begin{subfigure}[t]{0.235\textwidth}
        \includegraphics[pagebox=cropbox,clip,width=\linewidth]{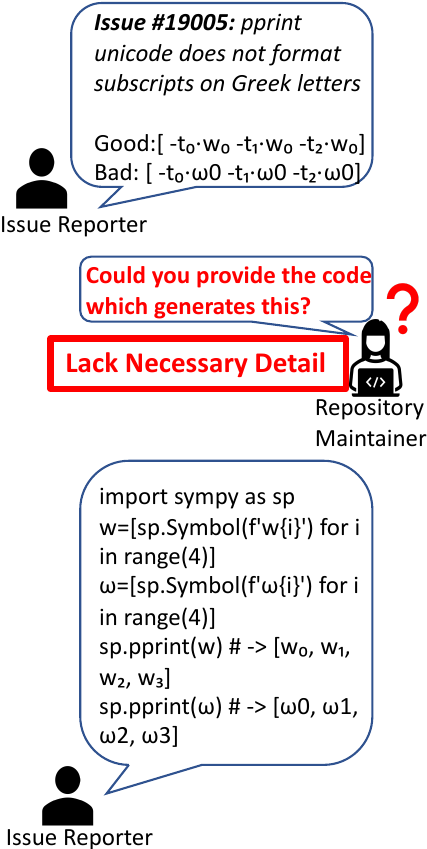}
        \caption{IS-1(sympy\_\_sympy-20916): The issue is too vague, confusing even the maintainer, so the model likely won’t understand either.}
        \label{fig:is_example}
      \end{subfigure}
      \Description{Four examples of PR-issue misalignment: a pull request that closes multiple issues, a defective pull request that causes random test failures, a follow-up pull request that addresses only the remaining bug, and an issue whose key requirements appear only in later discussion.}
      \caption{Examples of four misalignment patterns.}
      \label{fig:examples_4}
      \captionsetup{justification=centering} 
    \end{figure*}

    \textbf{Manual Examination Process.}\quad Following established                          
  practices in qualitative SE research~\cite{groundedref1,                               
  groundedref2}, we apply open coding~\cite{opencoding} to derive a
  misalignment taxonomy (RQ1). First, we collect per instance: issue-side
  artifacts (description, discussion), PR-side artifacts (description,                     
  discussion, code review, commits, changed files), and cross-reference
  metadata. Then, we code all 500 instances, iteratively grouping codes                     
  into higher-level categories to form a draft taxonomy; a single                        
  instance may exhibit multiple patterns. Last, the draft taxonomy was                           
  reviewed by the second author; disagreements were resolved with a
  third author until consensus. 

  \textbf{Leaderboard Data Collection.}\quad To assess the impact of
  misalignment on evaluation reliability, we collect per-instance                    
  resolution data from the official SWE-bench experiments                                
  repository~\cite{sweexperiments}, recording per-instance resolution counts for 131 leaderboard agents. 

    \subsection{Results and Analysis}
    \label{sec:study_results}

    \subsubsection{RQ1: Taxonomy and Prevalence of Misalignment}
    \label{sec:rq1}

    Table~\ref{tab:misalignment_patterns} summarizes the taxonomy of five                    
  patterns with 11 fine-grained scenarios, with instance counts in parentheses. \rev{The five patterns correspond to distinct root causes in the PR--Issue pairing pipeline. They are orthogonal by design: an instance may carry multiple labels when independent failure modes co-occur, while each label still denotes a separate reason for misalignment. This keeps counts interpretable as pattern-level evidence rather than catch-all defects.} Our classification also follows the \textit{text-driven, code-validation} principle (Section ~\ref{sec:intro}). We elaborate on each pattern below. 

  \textbf{PR Scope Creep (SC)} occurs when the PR delivers functionality                   
  beyond what the issue requests. We consider this a misalignment                          
  because it creates a mismatch between the task and its evaluation: the                   
  test patch validates requirements absent from the                                        
  \texttt{problem\_statement}, penalizing agents that correctly solve                      
  the stated problem. Specifically, SC-3 refers to pre-existing bugs in the repository, whereas SC-4 refers to patches addressing unrelated Issues.
  
  \emph{Example.} Figure~\ref{fig:sc1_example} shows
  \texttt{sphinx-doc\_\_sphinx-10614}: the \texttt{pro\allowbreak blem\_statement}
  captures only issue \texttt{\#10570}, but the PR closes                     
  three distinct issues. An agent correctly resolving \texttt{\#10570} may still
  fail if it does not coincidentally address the two unstated issues.

    \textbf{Defective PR (DP)} captures cases where the PR itself is flawed---introducing new bugs or providing an incomplete fix which requires subsequent corrective work. We consider this a type of misalignment because the test patch may encode buggy behavior as the expected output or validate only a partial solution.

    \emph{Example.} Figure~\ref{fig:dp_example} shows \texttt{sympy\_\_sympy-13852}. The PR fixes the reported issue but introduces random test failures, requiring a follow‑up fix. Yet the benchmark treats the original defective PR as ground truth.

\textbf{Follow-up PR (FP)} is the converse of DP: the PR supplements
  or fixes a previous PR for the same issue. We consider this a                         
  misalignment because the codebase already contains the earlier PR's                      
  changes, so the model benefits from prior work and the evaluation does
  not reflect its true capability. 

  \emph{Example.} Figure~\ref{fig:fp_example} shows                                        
  \texttt{django\_\_django\-12125}: a previous PR partially fixed the                      
  inner-class path issue for \texttt{Field} subclasses, but                                
  \texttt{Enum} serialization remained broken. The current PR addresses                  
  only that remaining gap; the codebase already contains the earlier                   
  fix.

\textbf{Incomplete Specification (IS)} occurs when the original issue
  description is supplemented or revised in later discussion. We
  consider this a misalignment because agents receive only the initial
  description, which in such cases lacks information needed for
  resolution. Following the standard bug report format, which includes
\textit{code to reproduce, actual behavior, and expected
behavior}~\cite{bettenburg_what_2008,issuetemp1,issuetemp2,issuetemp3,issuetemp4},
we distinguish two forms of incomplete specification, both grounded
in maintainer consensus. ~(1)The issue is \textit{incomplete} when a
maintainer explicitly requests a missing component. This type of misalignment resembles the "unspecified" criterion in SWE-bench-verified~\cite{annotation_instruction}, but is more objective: insufficiency is only flagged when the repository maintainer also fails to understand the issue. \rev{For example, an issue specified mainly through an external link is valid if developers raise no clarification request, but misaligned once such a request appears.} We adopt this standard because if a maintainer, a human expert familiar with the project, cannot infer the requirements from the description alone, a language model would likely struggle to do so as well. ~(2) The issue is
\textit{redefined} when a component is revised by maintainer
agreement (\eg the expected behavior). Unlike the incomplete case, here the original description does provide necessary components, but one or more of them are superseded through maintainer discussion. An agent following the original description would therefore target an outdated requirement. 
  
  \emph{Example.} Figure~\ref{fig:is_example} shows
  \texttt{sympy\_\_sympy-20916}: the issue describes incorrect
  formatting and expected output, but lacks necessary detail. At a
  maintainer's request, the reporter supplies code containing
  key information absent from the \texttt{problem\_statement}.

    \textbf{Unspecified Literal (UL)} occurs when the test patch asserts exact literal values (\eg exception messages) introduced by the PR but absent from the issue. We consider this a type of misalignment because it penalizes agents that produce semantically correct outputs differing in literal form. This reflects an overly strict test oracle that requires implementation-specific details unrecoverable from the issue description alone.

    \emph{Example.} \texttt{astropy\_\_astropy-13033}: the issue requests ``\textit{an exception that tells the user which required columns are actually missing}'' without specifying the exact wording. However, the test patch asserts the precise string \texttt{``TimeSeries object is invalid -- expected {[}'time',\,'a'{]} as the first columns but found {[}'time',\,'b'{]}''}, rejecting correct fixes with different wording.

    \finding{13.6\% (68/500) of SWE-bench Verified instances exhibit PR-Issue misalignment. Defective PR (44.1\%), PR Scope Creep (32.4\%), and Incomplete Specification (26.5\%) are the three most prevalent pattern families.}

    \subsubsection{RQ2: Impact on Evaluation Reliability}
    \label{sec:rq2}
    In this RQ, we investigate the practical impact of PR-Issue misalignment. \rev{We correlate misalignment labels with resolution counts across 131 leaderboard agents and recompute the leaderboard~\cite{leaderboard} after excluding misaligned instances.}

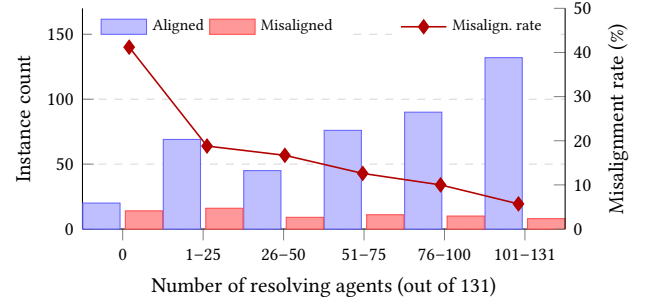
\begin{figure}[!t]
\centering
\begin{tikzpicture}
\begin{axis}[
    width=0.94\columnwidth,
    height=4.5cm,
    ybar,
    bar width=14pt,
    xlabel={Number of resolving agents (out of 131)},
    ylabel={Instance count},
    ymin=0, ymax=170,
    xtick=data,
    xticklabels={0, 1--25, 26--50, 51--75, 76--100, 101--131},
    x tick label style={font=\footnotesize},
    y tick label style={font=\footnotesize},
    label style={font=\small},
    axis y line*=left,
    axis x line*=bottom,
    ymajorgrids=true,
    grid style={dashed, gray!30},
    legend style={at={(0.02,1)}, anchor=north west, font=\scriptsize, draw=none,legend columns=-1},
    area legend,
]
\addplot[fill=blue!25, draw=blue!60] coordinates {
    (1,20) (2,69) (3,45) (4,76) (5,90) (6,132)
};
\addlegendentry{Aligned}

\addplot[fill=red!40, draw=red!70] coordinates {
    (1,14) (2,16) (3,9) (4,11) (5,10) (6,8)
};
\addlegendentry{Misaligned}
\end{axis}

\begin{axis}[
    width=0.94\columnwidth,
    height=4.5cm,
    axis y line*=right,
    axis x line=none,
    ylabel={Misalignment rate (\%)},
    ylabel style={font=\small},
    ymin=0, ymax=50,
    ytick={0,10,20,30,40,50},
    y tick label style={font=\footnotesize},
    xtick=data,
    xticklabels={},
    xmin=0.4, xmax=6.6,
    legend style={at={(0.98,1)}, anchor=north east, font=\scriptsize, draw=none},
]
\addplot[color=red!70!black, mark=diamond*, semithick, mark size=2.5pt] coordinates {
    (1,41.2) (2,18.8) (3,16.7) (4,12.6) (5,10.0) (6,5.7)
};
\addlegendentry{Misalign.\ rate}
\end{axis}
\end{tikzpicture}
\Description{A bar and line chart showing that the misalignment rate declines monotonically from 41.2 percent for instances solved by no agents to 5.7 percent for instances solved by more than 100 agents.}
\caption{Misalignment rate decreases monotonically as instances are resolved by more agents. Bars show instance counts (left axis); the line shows the misalignment rate within each bucket (right axis). Among the 34 never-resolved instances, 41.2\% are misaligned, compared to only 5.7\% among instances solved by $>$100 agents.}
\label{fig:rq2_trend}
\end{figure}

    \textbf{Misalignment--Resolution Correlation.}\quad \textit{(1) Methodology.} We bucket the 500 instances by the number of resolving agents and compute each bucket's misalignment rate (Figure~\ref{fig:rq2_trend}). \textit{(2) Results and Discussion.} Among the 500 instances, 34 have never been resolved by any of the
    131 leaderboard agents. Of these, 14 (41.2\%) exhibit PR-Issue
    misalignment, suggesting that the hardest instances \rev{may correlate with}
    agents due to unstated evaluation requirements, not task complexity
    alone. More broadly, Figure~\ref{fig:rq2_trend} shows a monotonic
  decrease in misalignment rate as instances are resolved by more
  agents: from 41.2\% among never-resolved instances to 5.7\% for those
    resolved by over 100 agents. Well-aligned instances are solvable by
    many agents, whereas misaligned instances introduce extraneous
    difficulty that few agents overcome by coincidence. These findings are consistent with OpenAI's concurrent
  analysis~\cite{swebenchverifiednolonger}, though we arrive at a
  similar conclusion from a distinct perspective: systematic PR-Issue
  misalignment in benchmark construction.

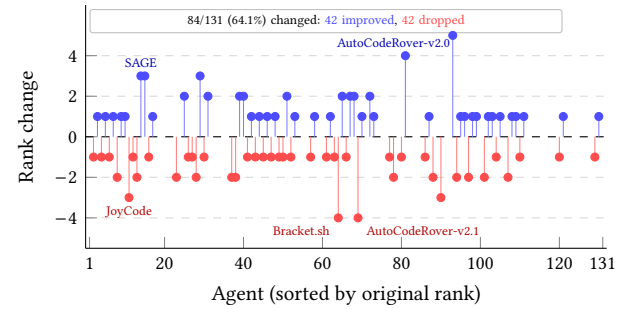
\begin{figure}[!t]
\centering
\begin{tikzpicture}
\begin{axis}[
    width=\columnwidth,
    height=4.8cm,
    xlabel={Agent (sorted by original rank)},
    ylabel={Rank change},
    ymin=-5.5, ymax=6.5,
    xmin=0, xmax=132,
    xtick={1,20,40,60,80,100,120,131},
    ytick={-4,-2,0,2,4},
    x tick label style={font=\footnotesize},
    y tick label style={font=\footnotesize},
    label style={font=\small},
    ymajorgrids=true,
    grid style={dashed, gray!30},
    axis y line*=left,
    axis x line*=bottom,
    clip=false,
]
\addplot[black, thin, dashed] coordinates {(0,0) (132,0)};
\addplot[only marks, mark=*, mark size=1.5pt, blue!70] coordinates {
    (3,1)(5,1)(7,1)(9,1)(10,1)(14,3)(15,3)(17,1)(25,2)(29,3)(31,2)
    (39,2)(40,2)(42,1)(44,1)(46,1)(48,1)(51,2)(53,1)(58,1)(62,1)
    (65,2)(67,2)(68,2)(70,1)(72,2)(73,1)(81,4)(87,1)(93,5)(95,1)
    (96,1)(98,1)(99,1)(102,1)(103,1)(105,1)(108,1)(109,1)(111,1)
    (121,1)(130,1)
};
\addplot[only marks, mark=*, mark size=1.5pt, red!70] coordinates {
    (2,-1)(4,-1)(6,-1)(8,-2)(11,-3)(12,-1)(13,-2)(16,-1)(23,-2)
    (26,-1)(27,-1)(28,-2)(30,-1)(37,-2)(38,-2)(41,-1)(43,-1)(45,-1)
    (47,-1)(49,-1)(50,-1)(52,-1)(57,-1)(61,-1)(63,-1)(64,-4)(66,-1)
    (69,-4)(77,-1)(78,-2)(80,-1)(86,-1)(88,-2)(90,-3)(94,-2)(97,-2)
    (101,-2)(104,-1)(107,-2)(110,-1)(120,-1)(129,-1)
};
\addplot[ycomb, blue!50, thin] coordinates {
    (3,1)(5,1)(7,1)(9,1)(10,1)(14,3)(15,3)(17,1)(25,2)(29,3)(31,2)
    (39,2)(40,2)(42,1)(44,1)(46,1)(48,1)(51,2)(53,1)(58,1)(62,1)
    (65,2)(67,2)(68,2)(70,1)(72,2)(73,1)(81,4)(87,1)(93,5)(95,1)
    (96,1)(98,1)(99,1)(102,1)(103,1)(105,1)(108,1)(109,1)(111,1)
    (121,1)(130,1)
};
\addplot[ycomb, red!50, thin] coordinates {
    (2,-1)(4,-1)(6,-1)(8,-2)(11,-3)(12,-1)(13,-2)(16,-1)(23,-2)
    (26,-1)(27,-1)(28,-2)(30,-1)(37,-2)(38,-2)(41,-1)(43,-1)(45,-1)
    (47,-1)(49,-1)(50,-1)(52,-1)(57,-1)(61,-1)(63,-1)(64,-4)(66,-1)
    (69,-4)(77,-1)(78,-2)(80,-1)(86,-1)(88,-2)(90,-3)(94,-2)(97,-2)
    (101,-2)(104,-1)(107,-2)(110,-1)(120,-1)(129,-1)
};
\node[above, font=\tiny, blue!70!black] at (axis cs:93,5) {ugaiforge};
\node[above, font=\tiny, blue!70!black] at (axis cs:78,4) {AutoCodeRover-v2.0};
\node[below left, font=\tiny, red!70!black] at (axis cs:64,-4) {Bracket.sh};
\node[below right, font=\tiny, red!70!black] at (axis cs:69,-4) {AutoCodeRover-v2.1};
\node[above, font=\tiny, blue!70!black] at (axis cs:14,3) {SAGE};
\node[below, font=\tiny, red!70!black] at (axis cs:11,-3) {JoyCode};
\node[anchor=north west, font=\tiny, text width=0.72\columnwidth, align=center,
      fill=white, inner sep=2pt, draw=gray!50, rounded corners=1pt]
      at (axis cs:1,6.3) {84/131 (64.1\%) changed: \textcolor{blue!70}{42 improved}, \textcolor{red!70}{42 dropped}};
\end{axis}
\end{tikzpicture}
\Description{A lollipop chart of rank changes for 131 agents after removing 68 misaligned instances. Eighty-four agents change rank: 42 improve and 42 drop, with changes ranging from minus four to plus five positions.}
\caption{\rev{Per-agent rank change after excluding 68 misaligned instances, ordered by original rank. Each stem shows the magnitude and direction of rank movement (\textcolor{blue!70}{blue}: improved; \textcolor{red!70}{red}: dropped).}}
\label{fig:rerank_lollipop}
\end{figure}

  \rev{\textbf{Leaderboard Re-ranking.}\quad 
  \textit{(1) Methodology.} Using official resolution data~\cite{sweexperiments}, we recompute each agent's pass rate on the 432 aligned instances after excluding the 68 misaligned ones, then re-rank agents by the new pass rate.
  \textit{(2) Results and Discussion.} Figure~\ref{fig:rerank_lollipop} shows that 84 of 131 agents (64.1\%) change rank. Most shifts are within 1--2 ranks, but outliers reach $+5$ (ugaiforge) and $-4$ (Bracket.sh, AutoCodeRover-v2.1), and 9 of the top 10 positions change. Pass rates rise by 2.77 pp on average, with gains ranging from $-0.24$ to $+4.49$ pp, indicating uneven reliance on misaligned instances.}

    \begin{figure*}[t]
    \vspace{2mm}
    \centering
    \includegraphics[pagebox=cropbox,clip,width=\textwidth]{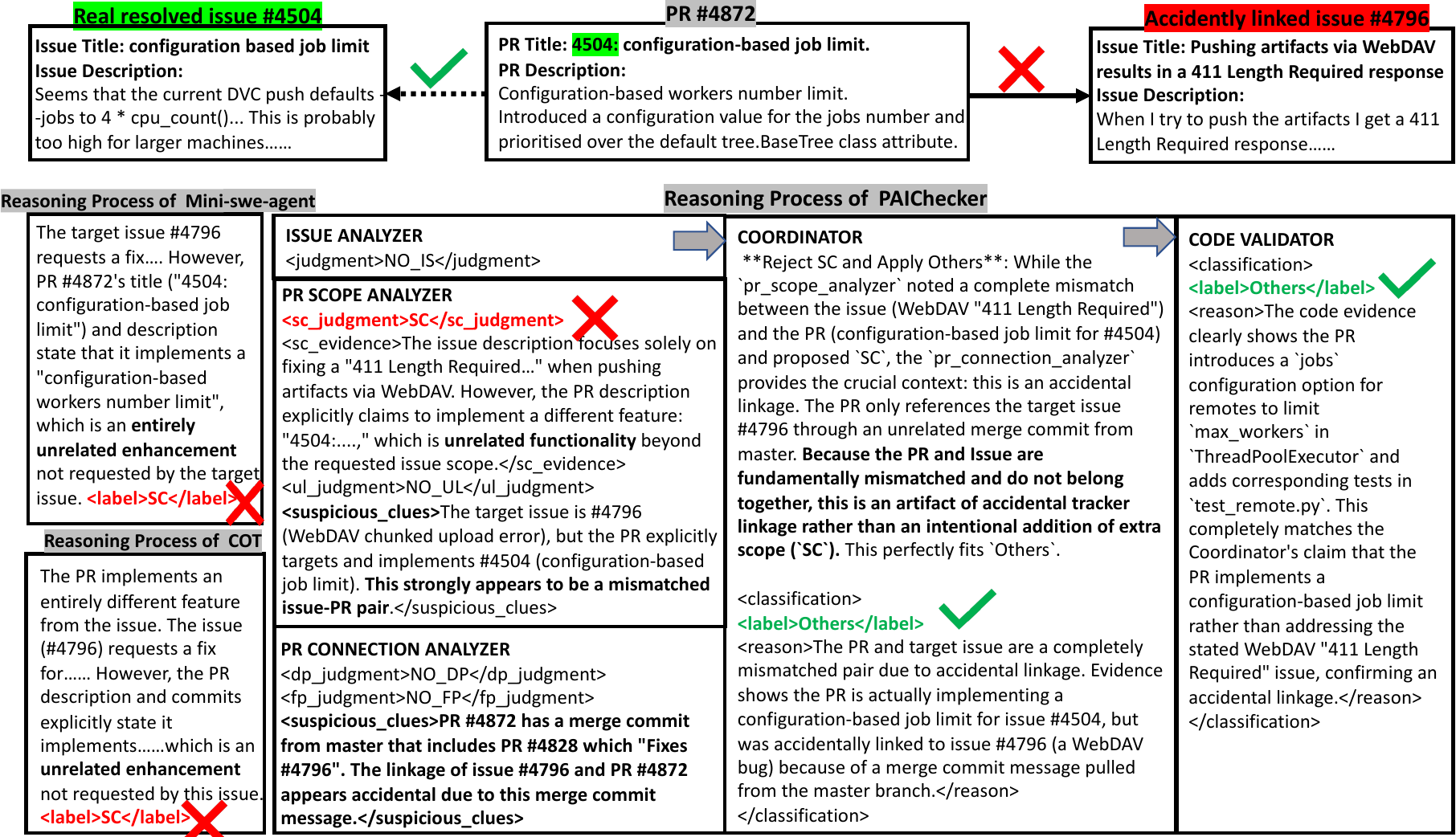}
    \Description{A motivating example in which a pull request is accidentally linked to one issue but fixes another. Two baselines force the example into a predefined class, whereas PAIChecker combines subagent clues and assigns the catch-all Others label.}
    \caption{Motivating Example \texttt{iterative\_\_dvc-4872}: PR~\#4872 is accidentally
    linked to issue~\#4796 but actually fixes issue~\#4504. Backbone:
    Gemini-3.1-Pro Preview. Mini-SWE-Agent and CoT force-classify as SC;
    \tool's coordinator correctly drops SC and assigns \textit{Others} by
    synthesizing suspicious clues from two subagents.}
    \label{fig:motivation}
    \end{figure*}

\finding{\rev{Misalignment rate decreases from 41.2\% to 5.7\% as resolution count rises, and excluding 68 misaligned instances changes 64.1\% of agent rankings. This correlation is not causal evidence, but shows that construction defects add extraneous difficulty and distort standings unevenly.}}
    
    \subsection{Motivating Examples}
    \label{sec:motivation}

    In this subsection, we use concrete examples to illustrate why general
    prompting and agentic approaches fall short for misalignment detection,
    motivating the design of \tool. We adapt Mini‑SWE‑Agent~\cite{minisweagent} as a representative tool-augmented agent by adjusting its prompt and equipping it with GitHub API Access, and include Chain-of-Thought (CoT)~\cite{cot} prompting as a representative prompting technique, both on Gemini‑3.1‑Pro Preview. We demonstrate limitations common to these                                                 
  monolithic approaches.  

We observe a monolithic approach struggles when all artifacts and detection rules are packed into a single prompt, because each misalignment pattern demands a distinct reasoning workflow over a different artifact subset: SC compares issue scope against PR-side claims, DP traces cross-PR relationships, and IS tracks specification changes in discussion threads. A single prompt cannot specialize in all simultaneously: detection logic for one pattern interferes with another, diluting model's focus; and artifacts critical to one pattern, such as cross-issue references for DP, are overshadowed by more voluminous ones. \textit{This motivates Phase I of \tool: decomposing detection into three dedicated subagents, each with a focused artifact subset and workflow tailored to its target pattern.}

We further examine a beyond-taxonomy case in
  Figure~\ref{fig:motivation}, where a PR and its linked issue are
  fundamentally unrelated due to accidental tracker linkage. This
  example exposes two problems with monolithic approaches. First, both
  Mini-SWE-Agent and CoT fail to assign the \textit{Others} label
  despite having the option, defaulting instead to SC, the nearest
  predefined category. \textit{Others} requires concluding that no
  predefined pattern applies, a harder judgment than matching any
  single one. \textit{This presents generalizability challenge and motivates Phase~II of \tool: a
  coordinator that collects suspicious clues from subagents before label
  commitment, allowing beyond-taxonomy signals to surface explicitly.}
  Second, both approaches incorrectly assign SC, creating a
  reasoning--label inconsistency: their reasoning describes the PR as
  ``entirely unrelated,'' which directly contradicts SC. Similar inconsistencies appear in other cases (\eg
  \texttt{MONAI-6793}), where cross-issue references are mistaken for scope creep despite no actual
  scope expansion. Once committed in a single pass, such errors
  cannot be revisited. \textit{This presents textual-level inconsistency challenge and motivates the coordinator to also
  re-verify each label against its cited evidence and drop inconsistent
  labels.}

    \begin{figure*}[t]
        \vspace{2mm}
        \centering
        \includegraphics[pagebox=cropbox,clip,width=0.9\textwidth]{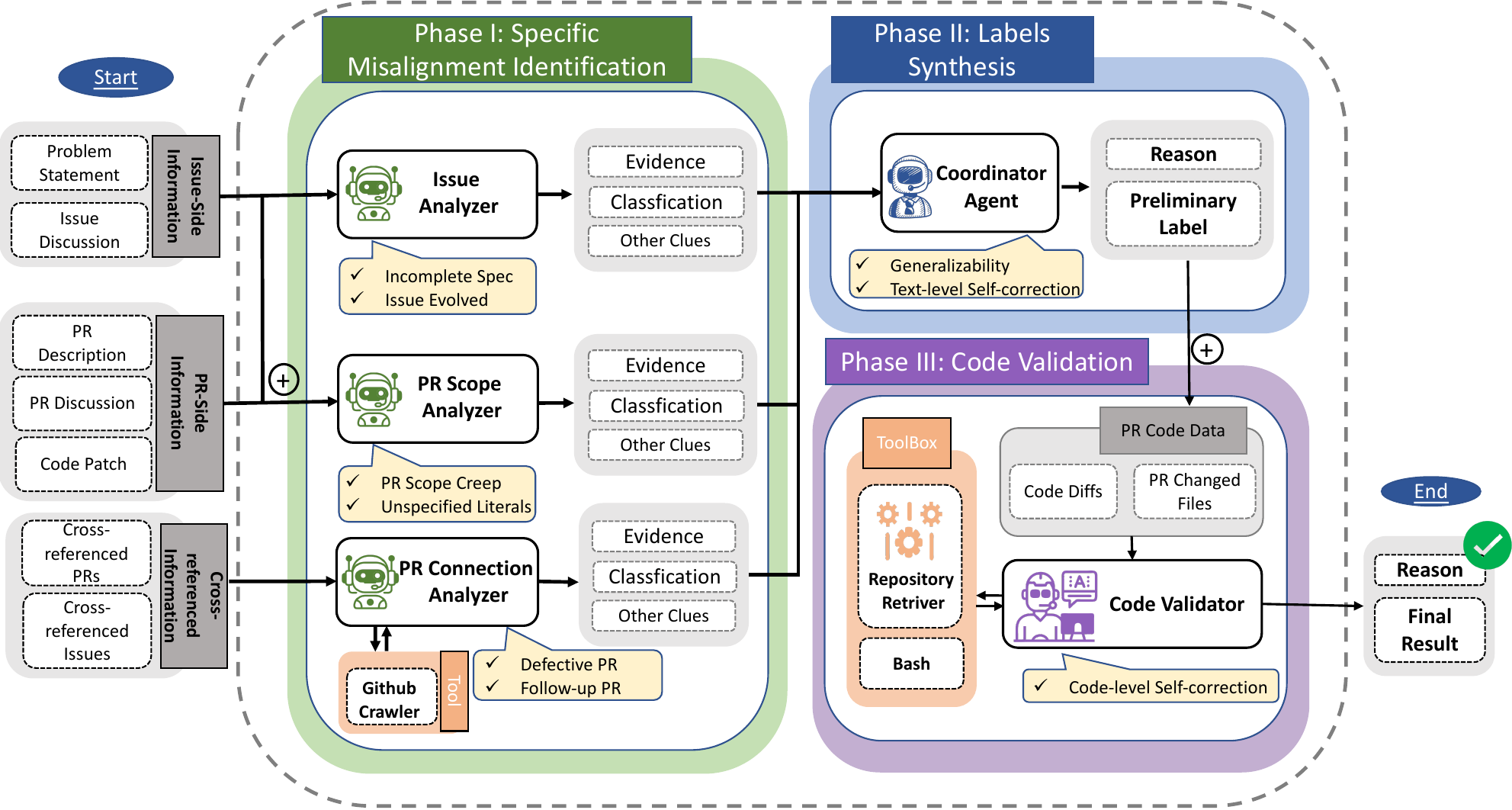}
        \Description{The three-phase PAIChecker workflow. Three specialized agents analyze issue, pull-request scope, and pull-request connections; a coordinator synthesizes preliminary labels; and a code validator checks the labels against code and supplementary context.}
        \caption{Workflow of \tool. Phase~I performs specific misalignment identification through three specialized subagents. Phase~II synthesizes their outputs into a preliminary label and rationale. Phase~III validates the preliminary result against code-level evidence and supplementary GitHub context to produce the final decision.}
        \label{fig:overview}
    \end{figure*}

Following the \textit{text-driven, code-validation} principle
  (Section~\ref{sec:rq1}), code-level verification is necessary to
  catch false positives that textual analysis alone cannot catch. For
  instance, a PR claiming to fix multiple issues may actually address
  duplicates sharing one root cause, or claimed extra functionality may
  not materialize in the code diff. \textit{This presents code-level inconsistency challenge and motivates Phase~III of \tool: a code validator that cross-references textual claims against
  implementation evidence.} Together with the coordinator's text-level
  re-verification, this forms a \textit{two-level self-correction
  mechanism} to ensure the accuracy of final output.  
    
    \section{Methodology: \tool}
    \label{sec:method}

    In this section, we present \tool, a multi-agent framework for detecting and categorizing PR-Issue
  misalignment. As shown in Figure~\ref{fig:overview}, \tool has three phases: Phase~I performs pattern-specific identification with three focused subagents, Phase~II synthesizes their reports into preliminary
  labels and can assign \textit{Others}, and Phase~III validates the textual judgment against code-level
  evidence.

    Across all three phases, we enforce a separation of responsibility: the authority to assign predefined labels is exclusive to Phase~I; Phases~II and~III hold only veto power. This ensures that detection decisions are made in the focused context where relevant evidence is most concentrated, while subsequent phases serve as progressive self-correction filters.

    \subsection{Phase I: Specific Misalignment Identification}
    \label{sec:subagents}

    In this subsection, we describe Phase~I, which detects the five misalignment patterns using three parallel subagents, each focused on distinct reasoning: the Issue Analyzer targets Incomplete Specification~(IS), primarily drawing on issue-side artifacts; the PR Scope Analyzer targets PR Scope Creep~(SC) and Unspecified Literal~(UL) through issue-versus-PR scope comparison; and the PR Connection Analyzer targets Defective PR~(DP) and Follow-up PR~(FP) through temporal cross-PR reasoning. This grouping keeps each subagent's context compact and reasoning focused. All three produce a uniform structured output: (i)~a \textit{judgment} per target pattern, (ii)~supporting \textit{evidence} for text-level self-correction, and (iii)~\textit{suspicious clues}---anomalies outside the subagent's target patterns that support beyond-taxonomy generalization.

    \subsubsection{Issue Analyzer}
    \label{sec:issue_analyzer}

The Issue Analyzer operationalizes IS detection by comparing the                                                                         
  issue description against its discussion thread. As mentioned in Section~\ref{sec:rq1}: a well-formed bug                                                                     
  report comprises three key components: reproduction code, actual                                                                         
  behavior, and expected behavior. Based on this, the                                                                             
  subagent performs two checks. First, whether the issue is                                                                                
  incomplete: it identifies whether a maintainer explicitly requests                                                                       
  the reporter to supplement any of these components, grounding the                                                                        
  judgment in community consensus rather than subjective assessment.                                                                     
 Second, whether the issue has been redefined: it checks whether any of these components differs between the original description and the discussion; a change in any component signals that the addressed problem has shifted.       

    \subsubsection{PR Scope Analyzer}
    \label{sec:scope_analyzer}

    The PR Scope Analyzer detects SC and UL in a unified pass. Both                                                                          
  patterns require establishing the issue's requested scope as a                                                                         
  baseline and checking whether the PR exceeds it, so a single
  subagent establishes the baseline once and verifies both patterns                                                                        
  together. This subagent receives the issue description as scope
  baseline, PR-side text for what the PR claims to deliver, and both                                                                       
  patches for UL detection. 

    Following the text-driven, code-validation principle, SC is triggered only when PR-side textual artifacts explicitly claim work beyond the issue's scope. UL is triggered when all three conditions hold: (1) the test patch asserts a hardcoded literal (\eg a string, number, or exception message); (2) the literal is newly added in the code patch as a fixed constant, rather than read from existing metadata; and (3) it does not appear verbatim in the issue description.  

    \subsubsection{PR Connection Analyzer}
    \label{sec:connection_analyzer}

    The PR Connection Analyzer detects DP and FP, both involve temporal relationships between the current                                                                      PR and other PRs for the same issue. Tracing such cross-PR                                                                             dependency chains requires information beyond the current                                                                                instance's artifacts, so we equip this subagent with GitHub API
  access to actively search for and retrieve related PRs. 

    For DP detection, the agent searches for PRs merged after the current one that still address the same issue. Each candidate is verified for merge status to distinguish corrective follow-ups from unrelated references. For FP detection, the agent searches for earlier merged PRs that partially addressed the same issue, checking whether the current PR is a supplementary contribution rather than an independent fix, based on merge chronology and follow-up language in PR descriptions.

    \subsection{Phase II: Label Synthesis}
    \label{sec:coordinator}

    In this subsection, we describe Phase~II, which deploys a coordinator agent to synthesize the three subagents' outputs into a preliminary label set. The coordinator has two responsibilities: beyond-taxonomy generalization and text-level self-correction. In both roles, it holds only veto power
  over predefined labels: it can remove but not add any
  predefined label not assigned by Phase~I, because it lacks each subagent's specialized context.

  For beyond-taxonomy generalization, the coordinator collects
  suspicious clues from all three subagents and evaluates whether they
  collectively point to a misalignment outside the five predefined
  categories; when they do, it assigns the \textit{Others} label.
  This design decouples anomaly discovery from judgment: subagents
  surface anomalies within their focused scope, and the coordinator
  synthesizes all three evidence streams. For text-level
  self-correction, the coordinator independently assesses whether each
  judgment is adequately supported by its evidence; if the evidence is
  insufficient or self-contradictory, it drops the label, catching
  over-detections before they reach code validation.

    \subsection{Phase III: Code Validation}
    \label{sec:code_validator}

    In this subsection, we describe Phase~III, which deploys a code validator agent responsible for code-level self-correction. The preceding phases make detection decisions from textual evidence;
  the code validator cross-references these decisions against the                                                                          
  actual implementation to filter false positives that textual
  reasoning alone cannot catch. It can only                                                                          
  remove labels, not add them.                                                                                                             
                                                                                                                                           
  To perform this verification, the code validator is augmented with                                                                       
  repository-level file retrieval capability, enabling it to download                                                                      
  full source files and inspect changed files in their complete                                                                          
  context. This is necessary because the \texttt{patch} contains only
  changed hunks, and determining whether a claimed misalignment
  materializes often requires inspecting surrounding code.



 \section{Evaluation Design}
    \label{sec:eval_design}
        This section details the evaluation setup for \tool. We aim to address the following research questions:
    
    \begin{itemize}[leftmargin=*]
        \item \textbf{RQ3 (Effectiveness):} How effective is \tool in detecting
            and categorizing PR-Issue misalignment?
        \item \textbf{RQ4 (Ablation):} How does each component of \tool
            contribute to the overall detection performance?
        \item \textbf{RQ5 (Label Changes):} How and why do labels change across the three pipeline phases?
    \end{itemize}       
    \label{sec:eval_dataset}

        \begin{table}[!t]
        \centering
        \small
        \caption{Statistics of the evaluation dataset. An instance may carry multiple labels.}
        \label{tab:eval_dataset}
        \resizebox{0.95\columnwidth}{!}{%
        \begin{tabular}{lcccc}
            \toprule
            \multirow{2}{*}{\textbf{Category}} &
            \multicolumn{2}{c}{\textbf{SWE-Gym}} &
            \multicolumn{2}{c}{\rev{\textbf{SWE-bench Multilingual}}} \\
            \cmidrule(lr){2-3} \cmidrule(lr){4-5}
            & \textbf{Count} & \textbf{Percentage} & \rev{\textbf{Count}} & \rev{\textbf{Percentage}} \\
            \midrule
            Total instances & 2{,}438 & 100\% & \rev{300} & \rev{100\%} \\
            Aligned & 1{,}358 & 55.7\% & \rev{227} & \rev{75.7\%} \\
            Misaligned & 1{,}080 & 44.3\% & \rev{73} & \rev{24.3\%} \\
            \midrule
            \quad SC & 428 & 17.6\% & \rev{33} & \rev{11.0\%} \\
            \quad DP & 150 & 6.2\% & \rev{24} & \rev{8.0\%} \\
            \quad IS & 341 & 14.0\% & \rev{13} & \rev{4.3\%} \\
            \quad UL & 387 & 15.9\% & \rev{9} & \rev{3.0\%} \\
            \quad FP & 109 & 4.5\% & \rev{2} & \rev{0.7\%} \\
            \quad Others & 16 & 0.7\% & \rev{1} & \rev{0.3\%} \\
            \bottomrule
        \end{tabular}
        }
    \end{table}

\subsection{Evaluation Dataset}
  \rev{We evaluate on two datasets disjoint from preliminary study: SWE-Gym~\cite{swegym} (2,438 Python tasks) and SWE-bench Multilingual~\cite{swebench} (300 tasks covering C, C++, Go, Java, JavaScript/TypeScript, PHP, Ruby, and Rust). Selection followed five criteria: (i) SWE-bench-style automated PR--Issue pairing; (ii) no manual specification rewriting, so raw defects remain observable; (iii) feasible for full manual annotation yet sufficiently large; (iv) multi-repo and widely adopted; and (v) PR identifiers available for full-content access.} Two annotators with development experience independently labeled all instances following our taxonomy (Section~\ref{sec:rq1}), guided by structured questions adapted from SWE-bench annotation instructions~\cite{annotation_instruction}. They achieved Cohen's Kappa~\cite{kappa} of $\kappa = 0.91$ on binary judgment and $\kappa = 0.86$ on fine-grained labeling; disagreements were resolved by a third author. Table~\ref{tab:eval_dataset} summarizes the dataset statistics.

\rev{Misalignment rates differ across datasets: 13.6\% in SWE-bench Verified
  (Table~\ref{tab:misalignment_patterns}), 44.3\% in SWE-Gym, and 24.3\% in SWE-bench Multilingual, largely
  reflecting SWE-bench Verified's human curation. Patterns tied to underspecification and test coverage
  drop most sharply in Verified: UL decreases from 15.9\% (SWE-Gym) and 3.0\% (Multilingual) to 0.2\%, IS
  from 14.0\% and 4.3\% to 3.6\%, and SC from 17.6\% and 11.0\% to 4.4\%. In contrast, deeper PR--Issue
  relationship defects remain comparable: DP stays near 6.0--8.0\%, and FP remains low across all three
  datasets (0.6\%, 4.5\%, 0.7\%). These defects are harder to infer from the task alone and thus more
  likely to escape human curation.}

    \subsection{Compared Techniques}
    \label{sec:baselines}
We compare \tool against three typical prompting techniques and \rev{four} agent \rev{approaches}:~(1)\textit{Zero-Shot
  Prompting}~\cite{zeroshot_fewshot} provides the LLM with the task                                                               
  description, taxonomy definitions, and instance artifacts for direct                                                                     
  classification. ~(2)\textit{Few-Shot Prompting}~\cite{fewshot} augments                                                                     
  the zero-shot prompt with annotated examples from SWE-bench                                                                              
  Verified. ~(3)\textit{Chain-of-Thought Prompting(CoT)}~\cite{cot} further                                                                         
  adds explicit reasoning steps. ~(4)\textit{Mini-SWE-Agent}~\cite{minisweagent}                                                               
  is a general-purpose agent baseline: we change the default \textit{Bash} tool to\textit{ Curl }and provide the same instructions and artifact context as                                                                           
  \tool to ensure a fair comparison. \rev{(5)\textit{OpenHands}~\cite{openhands} is an agent-framework baseline evaluated with the same artifact context. (6) \textit{Claude Code}~\cite{claudecode} and (7) \textit{Codex}~\cite{codex} with their officially supported models.}

    \subsection{Evaluation Metrics}
    \label{sec:metrics}

   We evaluate PR-Issue misalignment detection from two complementary perspectives: (1)~a \textbf{binary classification} task that judges whether an instance is aligned or misaligned, and (2)~a \textbf{multi-label classification} task that identifies the specific misalignment patterns, where an instance may exhibit multiple patterns simultaneously. Accordingly, we adopt the following metrics:

    \textbf{Binary Detection.}\quad We measure accuracy, precision, recall, and F1 for the binary detection task.

    \textbf{Multi-Class Categorization.}\quad Exact Match (EM) serves as the primary metric, requiring the predicted label set to exactly match the ground truth. We supplement EM with macro‑averaged accuracy, precision, recall, and F1. For per‑label analysis, we report only F1 due to space constraints; F1 provides a balanced measure of precision and recall, making it sufficient for diagnostic purposes.

    \subsection{Backbone LLMs}
    \label{sec:backbone}

    We evaluate \tool and baselines across four state-of-the-art LLM backbones, including three closed-source models GPT-5.3-Codex~\cite{gpt53codex}, Claude-Sonnet-4.6~\cite{claudesonnet46}, Gemini-3.1-Pro-Preview~\cite{gemini31pro}, and one open-sourced model Qwen-3.5-Plus~\cite{qwen35}. All models were accessed through their respective APIs.


\begin{table}[!t]
\centering
\scriptsize
\caption{Binary misalignment detection performance (\%). Accuracy, Precision, Recall, and F1 are computed for the misaligned class; all differences are significant at $p < 0.001$. GPT: GPT-5.3-Codex; Claude: Claude-Sonnet-4.6; Gemini: Gemini-3.1-Pro-Preview; Qwen: Qwen-3.5-Plus. ZS: Zero-Shot; FS: Few-Shot; MSA: Mini-SWE-Agent; OH: OpenHands; CC: Claude Code; CX: Codex.}
\label{tab:rq3_binary}
\renewcommand{\arraystretch}{0.72}
\setlength{\tabcolsep}{2pt}
\resizebox{\columnwidth}{!}{%
\begin{tabular}{ll cccc cccc}
    \toprule
    & & \multicolumn{4}{c}{\textbf{SWE-Gym}} & \multicolumn{4}{c}{\rev{\textbf{Multilingual}}} \\
    \cmidrule(lr){3-6} \cmidrule(lr){7-10}
    \textbf{LLM} & \textbf{Meth.} & \textbf{BA} & \textbf{BP} & \textbf{BR} & \textbf{BF1} & \rev{\textbf{BA}} & \rev{\textbf{BP}} & \rev{\textbf{BR}} & \rev{\textbf{BF1}} \\
    \midrule
    \multirow{7}{*}{GPT} & ZS & 74.36 & 73.14 & 66.57 & 69.70 & \rev{79.67} & \rev{57.32} & \rev{64.38} & \rev{60.65} \\
    & FS & 75.72 & 75.31 & 67.22 & 71.04 & \rev{79.33} & \rev{57.33} & \rev{58.90} & \rev{58.11} \\
    & CoT & 75.72 & 73.33 & 71.02 & 72.15 & \rev{83.33} & \rev{68.25} & \rev{58.90} & \rev{63.24} \\
    & MSA & 72.48 & 83.04 & 47.59 & 60.51 & \rev{88.00} & \rev{84.91} & \rev{61.64} & \rev{71.43} \\
    & \rev{OH} & \rev{70.10} & \rev{79.59} & \rev{48.13} & \rev{59.99} & \rev{83.67} & \rev{83.33} & \rev{41.10} & \rev{55.05} \\
    & \rev{CX} & \rev{62.34} & \rev{61.18} & \rev{44.60} & \rev{51.59} & \rev{82.67} & \rev{81.82} & \rev{36.99} & \rev{50.94} \\
    & \textbf{Ours} & \textbf{88.11} & \textbf{83.47} & \textbf{91.20} & \textbf{87.17} & \rev{\textbf{91.00}} & \rev{\textbf{85.94}} & \rev{\textbf{75.34}} & \rev{\textbf{80.29}} \\
    \midrule
    \multirow{7}{*}{Claude} & ZS & 67.02 & 65.23 & 54.72 & 59.52 & \rev{56.33} & \rev{33.71} & \rev{82.19} & \rev{47.81} \\
    & FS & 69.03 & 68.24 & 56.30 & 61.69 & \rev{70.33} & \rev{43.94} & \rev{79.45} & \rev{56.59} \\
    & CoT & 73.54 & 68.20 & 75.46 & 71.65 & \rev{75.33} & \rev{49.54} & \rev{73.97} & \rev{59.34} \\
    & MSA & 82.36 & 81.13 & 78.43 & 79.76 & \rev{86.00} & \rev{70.13} & \rev{73.97} & \rev{72.00} \\
    & \rev{OH} & \rev{77.66} & \rev{76.82} & \rev{73.24} & \rev{74.99} & \rev{75.67} & \rev{50.00} & \rev{68.49} & \rev{57.80} \\
    & \rev{CC} & \rev{78.09} & \rev{75.30} & \rev{76.20} & \rev{75.75} & \rev{72.00} & \rev{43.82} & \rev{53.42} & \rev{48.15} \\
    & \textbf{Ours} & \textbf{87.49} & \textbf{82.76} & \textbf{90.65} & \textbf{86.52} & \rev{\textbf{91.67}} & \rev{\textbf{77.91}} & \rev{\textbf{91.78}} & \rev{\textbf{84.28}} \\
    \midrule
    \multirow{6}{*}{Gemini} & ZS & 74.36 & 67.93 & 79.81 & 73.39 & \rev{63.67} & \rev{37.32} & \rev{72.60} & \rev{49.30} \\
    & FS & 81.05 & 79.37 & 77.31 & 78.33 & \rev{77.67} & \rev{53.19} & \rev{68.49} & \rev{59.88} \\
    & CoT & 81.09 & 80.49 & 75.65 & 78.00 & \rev{83.33} & \rev{64.94} & \rev{68.49} & \rev{66.67} \\
    & MSA & 83.59 & 83.60 & 78.33 & 80.88 & \rev{59.33} & \rev{26.67} & \rev{38.36} & \rev{31.46} \\
    & \rev{OH} & \rev{79.65} & \rev{82.02} & \rev{69.26} & \rev{75.10} & \rev{86.67} & \rev{78.95} & \rev{61.64} & \rev{69.23} \\
    & \textbf{Ours} & \textbf{92.12} & \textbf{91.19} & \textbf{91.02} & \textbf{91.10} & \rev{\textbf{91.33}} & \rev{\textbf{88.52}} & \rev{\textbf{73.97}} & \rev{\textbf{80.60}} \\
    \midrule
    \multirow{6}{*}{Qwen} & ZS & 74.36 & 82.92 & 53.06 & 64.71 & \rev{83.00} & \rev{77.50} & \rev{42.47} & \rev{54.87} \\
    & FS & 73.75 & 80.73 & 53.52 & 64.37 & \rev{82.33} & \rev{76.32} & \rev{39.73} & \rev{52.25} \\
    & CoT & 72.76 & 84.90 & 46.85 & 60.38 & \rev{81.67} & \rev{76.47} & \rev{35.62} & \rev{48.60} \\
    & MSA & 74.77 & 77.65 & 60.46 & 67.99 & \rev{74.33} & \rev{45.00} & \rev{24.66} & \rev{31.86} \\
    & \rev{OH} & \rev{60.53} & \rev{65.97} & \rev{28.25} & \rev{39.56} & \rev{81.33} & \rev{\textbf{79.31}} & \rev{31.51} & \rev{45.10} \\
    & \textbf{Ours} & \textbf{80.56} & \textbf{90.51} & \textbf{62.69} & \textbf{74.07} & \rev{\textbf{85.33}} & \rev{78.43} & \rev{\textbf{54.79}} & \rev{\textbf{64.52}} \\
    \bottomrule
\end{tabular}%
}
\end{table}

    \section{Results and Analysis}
    \label{sec:results}

    \subsection{RQ3: Effectiveness of \tool}
    \label{sec:rq3}

    In this RQ, we evaluate the effectiveness of \tool from two complementary perspectives: \textit{binary misalignment detection}---whether an instance exhibits any misalignment (Section~\ref{sec:rq3_binary}), and \textit{multi-class categorization}---whether the specific misalignment patterns are correctly identified (Section~\ref{sec:rq3_multi}).


    \subsubsection{Effectiveness of Detecting Misalignment}
    \label{sec:rq3_binary}

Table~\ref{tab:rq3_binary} reports binary detection results.  \rev{\tool achieves the best Accuracy and F1 across all backbones  
  on both datasets. On SWE-Gym, it reaches 92.12\% Accuracy and 91.10\% F1 with Gemini; on SWE-bench    
  Multilingual, it reaches 91.67\% Accuracy and 84.28\% F1 with Claude.}
  
  \rev{On SWE-Gym, \tool improves Accuracy over the best baseline per backbone by 5.13--12.39 points: Mini-SWE-Agent is strongest on Claude/Gemini (82.36\%/83.59\%) but trails by 5.13/8.53 points, while GPT/Qwen gaps are 12.39/5.09 points. The advantage also holds on SWE-bench Multilingual, where \tool reaches 85.33--91.67\% Accuracy versus 83.00--88.00\% for the best baseline. Because Multilingual has fewer misaligned instances (24.3\% vs.\ 44.3\% in SWE-Gym), some baselines trade recall for precision; for example, OpenHands on Qwen has higher Precision than \tool (79.31\% vs.\ 78.43\%) but much lower Recall (31.51\%), F1 (45.10\% vs.\ 64.52\%), and Accuracy (81.33\% vs.\ 85.33\%).}

    \subsubsection{Effectiveness of Categorization}
    \label{sec:rq3_multi}

Table~\ref{tab:rq3_multi} shows multi-class categorization results. \rev{\tool achieves the best EM and Macro F1 across all backbones on both datasets. On SWE-Gym, EM ranges from 70.59\% (Qwen) to 84.66\% (Gemini), and Macro F1 from 58.98\% to 79.21\%; on SWE-bench Multilingual, Gemini reaches 90.67\% EM and 71.77\% Macro F1. We omit per-label F1 for FP and Others on Multilingual, as the 2 FP and 1 Others instances are insufficient for meaningful statistical comparison.}

\begin{table*}[!tbp]
\centering
\small
\renewcommand{\arraystretch}{0.67}
\caption{Multi-class categorization results (\%). EM requires the predicted label set to match the ground truth exactly; Macro metrics average all seven labels. Per-label columns report F1, and best values per backbone are in \textbf{bold}.}
\label{tab:rq3_multi}
\resizebox{0.98\textwidth}{!}{%
\begin{tabular}{ll c cccc ccccccc}
    \toprule
    & & & \multicolumn{4}{c}{\textbf{Macro}} & \multicolumn{7}{c}{\textbf{Per-Label F1}} \\
    \cmidrule(lr){4-7} \cmidrule(lr){8-14}
    \textbf{Backbone} & \textbf{Method} & \textbf{EM} & \textbf{Acc.} & \textbf{Prec.} & \textbf{Rec.} & \textbf{F1} & \textbf{SC} & \textbf{FP} & \textbf{DP} & \textbf{IS} & \textbf{UL} & \textbf{Others} & \textbf{No Mis.} \\
    \midrule
    \multicolumn{14}{l}{\rev{\textbf{SWE-Gym}}} \\
    \midrule
    \multirow{7}{*}{\shortstack[l]{GPT-5.3\\Codex}} & Zero-Shot & 58.20 & 88.40 & 49.36 & 41.26 & 41.01 & 64.70 & 63.60 & 37.30 & 38.80 & 4.90 & 0.00 & 77.80 \\
     & Few-Shot & 60.21 & 89.10 & 56.26 & 41.14 & 42.97 & 67.40 & 62.10 & 30.50 & 40.10 & 21.60 & 0.00 & 79.10 \\
     & Chain-of-Thought & 60.34 & 89.20 & 53.59 & 49.37 & 49.81 & 67.10 & 70.90 & 48.70 & 44.80 & 38.80 & 0.00 & 78.40 \\
     & Mini-SWE-Agent & 62.02 & 89.40 & 62.34 & 39.63 & 45.07 & 62.10 & 59.50 & 55.40 & 42.00 & 17.70 & 0.00 & 78.80 \\
     & \rev{OpenHands} & \rev{61.59} & \rev{90.00} & \rev{74.80} & \rev{39.27} & \rev{46.61} & \rev{54.16} & \rev{54.01} & \rev{11.85} & \rev{46.96} & \rev{59.22} & \rev{7.40} & \rev{79.37} \\
     & \rev{Codex} & \rev{54.36} & \rev{89.40} & \rev{71.28} & \rev{34.87} & \rev{43.32} & \rev{51.67} & \rev{46.02} & \rev{17.39} & \rev{46.80} & \rev{52.39} & \rev{6.90} & \rev{75.62} \\
     & \textbf{\tool} & \textbf{79.78} & \textbf{94.80} & \textbf{78.39} & \textbf{68.07} & \textbf{69.49} & \textbf{81.40} & \textbf{79.40} & \textbf{63.40} & \textbf{78.00} & \textbf{84.20} & \textbf{11.10} & \textbf{88.90} \\
    \midrule
    \multirow{7}{*}{\shortstack[l]{Claude-Sonnet\\4.6}} & Zero-Shot & 53.16 & 86.80 & 47.14 & 35.14 & 38.79 & 58.40 & 47.60 & 41.20 & 30.60 & 21.50 & 0.00 & 72.20 \\
     & Few-Shot & 55.58 & 87.70 & 49.77 & 39.70 & 43.29 & 63.10 & 61.20 & 43.50 & 29.10 & 24.70 & 7.40 & 74.00 \\
     & Chain-of-Thought & 57.47 & 88.50 & 57.83 & 50.79 & 53.09 & 65.60 & 63.30 & 50.20 & 58.70 & 37.90 & 20.70 & 75.20 \\
     & Mini-SWE-Agent & 65.26 & 91.50 & 67.07 & 55.97 & 60.34 & 61.00 & 62.00 & 60.00 & 57.40 & 71.90 & 26.70 & 83.40 \\
     & \rev{OpenHands} & \rev{65.88} & \rev{91.51} & \rev{67.80} & \rev{54.77} & \rev{58.81} & \rev{65.53} & \rev{68.26} & \rev{45.65} & \rev{63.32} & \rev{71.47} & \rev{15.38} & \rev{82.07} \\
     & \rev{Claude Code} & \rev{64.30} & \rev{90.74} & \rev{66.24} & \rev{56.93} & \rev{59.90} & \rev{61.68} & \rev{66.25} & \rev{52.38} & \rev{62.35} & \rev{66.38} & \rev{29.41} & \rev{80.84} \\
     & \textbf{\tool} & \textbf{76.62} & \textbf{94.20} & \textbf{71.64} & \textbf{74.39} & \textbf{72.79} & \textbf{79.60} & \textbf{77.10} & \textbf{72.30} & \textbf{76.10} & \textbf{79.70} & \textbf{36.40} & \textbf{88.30} \\
    \midrule
    \multirow{6}{*}{\shortstack[l]{Gemini-3.1\\Pro}} & Zero-Shot & 56.15 & 88.20 & 58.06 & 56.94 & 54.61 & 67.70 & 71.00 & 50.70 & 61.90 & 34.90 & 20.80 & 75.30 \\
     & Few-Shot & 66.28 & 91.20 & 64.70 & 59.79 & 59.31 & 66.80 & 63.50 & 53.90 & 61.10 & 65.40 & 21.30 & 83.20 \\
     & Chain-of-Thought & 66.20 & 91.20 & 65.74 & 58.06 & 60.19 & 68.90 & 67.00 & 52.70 & 63.70 & 54.80 & 30.80 & 83.40 \\
     & Mini-SWE-Agent & 69.40 & 92.40 & 69.16 & 57.10 & 61.36 & 61.10 & 68.20 & 73.20 & 61.40 & 70.90 & 9.10 & 85.60 \\
     & \rev{OpenHands} & \rev{67.50} & \rev{91.77} & \rev{69.60} & \rev{52.67} & \rev{57.56} & \rev{66.99} & \rev{76.84} & \rev{36.46} & \rev{63.30} & \rev{58.82} & \rev{16.22} & \rev{84.28} \\
     & \textbf{\tool} & \textbf{84.66} & \textbf{96.20} & \textbf{79.60} & \textbf{80.46} & \textbf{79.21} & \textbf{85.10} & \textbf{80.00} & \textbf{77.60} & \textbf{82.80} & \textbf{89.60} & \textbf{46.50} & \textbf{92.90} \\
    \midrule
    \multirow{6}{*}{\shortstack[l]{Qwen-3.5\\Plus}} & Zero-Shot & 62.02 & 89.60 & 67.86 & 43.04 & 49.04 & 61.50 & 58.50 & 44.40 & 53.50 & 13.50 & 25.00 & 79.90 \\
     & Few-Shot & 61.81 & 89.80 & 74.11 & 41.34 & 49.83 & 58.20 & 52.90 & 40.40 & 50.70 & 38.80 & 21.30 & 79.20 \\
     & Chain-of-Thought & 61.48 & 89.10 & 66.81 & 39.36 & 44.73 & 55.20 & 58.70 & 37.60 & 52.80 & 11.40 & 18.20 & 79.20 \\
     & Mini-SWE-Agent & 61.57 & 89.90 & 70.57 & 42.23 & 50.61 & 54.80 & 44.80 & 51.50 & 47.60 & 51.90 & 25.00 & 78.70 \\
     & \rev{OpenHands} & \rev{54.37} & \rev{87.81} & \rev{63.79} & \rev{25.77} & \rev{31.24} & \rev{48.11} & \rev{30.89} & \rev{23.23} & \rev{28.15} & \rev{13.79} & \rev{0.00} & \rev{74.51} \\
     & \textbf{\tool} & \textbf{70.59} & \textbf{92.05} & \textbf{75.90} & \textbf{52.24} & \textbf{58.98} & \textbf{74.20} & \textbf{68.50} & \textbf{52.20} & \textbf{54.00} & \textbf{52.40} & \textbf{27.30} & \textbf{84.40} \\
    \midrule
    \multicolumn{14}{l}{\rev{\textbf{SWE-bench Multilingual}}} \\
    \midrule
    \rev{\multirow{7}{*}{\shortstack[l]{GPT-5.3\\Codex}}} & \rev{Zero-Shot} & \rev{73.00} & \rev{92.38} & \rev{41.13} & \rev{34.23} & \rev{30.44} & \rev{67.57} & \rev{-} & \rev{15.38} & \rev{18.18} & \rev{25.64} & \rev{-} & \rev{86.29} \\
     & \rev{Few-Shot} & \rev{75.00} & \rev{92.86} & \rev{41.19} & \rev{35.77} & \rev{31.14} & \rev{72.00} & \rev{-} & \rev{15.38} & \rev{10.00} & \rev{34.29} & \rev{-} & \rev{86.28} \\
     & \rev{Chain-of-Thought} & \rev{79.67} & \rev{94.14} & \rev{47.14} & \rev{38.56} & \rev{35.43} & \rev{77.78} & \rev{-} & \rev{8.00} & \rev{28.57} & \rev{44.44} & \rev{-} & \rev{89.22} \\
     & \rev{Mini-SWE-Agent} & \rev{83.33} & \rev{95.33} & \rev{49.04} & \rev{35.91} & \rev{38.69} & \rev{79.41} & \rev{-} & \rev{27.59} & \rev{28.57} & \rev{42.86} & \rev{-} & \rev{92.41} \\
     & \rev{OpenHands} & \rev{81.00} & \rev{94.43} & \rev{53.97} & \rev{32.43} & \rev{35.77} & \rev{66.67} & \rev{-} & \rev{8.00} & \rev{19.05} & \rev{66.67} & \rev{-} & \rev{90.02} \\
     & \rev{Codex} & \rev{80.67} & \rev{94.14} & \rev{57.50} & \rev{30.15} & \rev{34.23} & \rev{58.82} & \rev{-} & \rev{8.00} & \rev{33.33} & \rev{50.00} & \rev{-} & \rev{89.47} \\
     & \rev{\textbf{\tool}} & \rev{\textbf{88.00}} & \rev{\textbf{96.33}} & \rev{\textbf{82.22}} & \rev{\textbf{70.91}} & \rev{\textbf{71.85}} & \rev{\textbf{82.19}} & \rev{-} & \rev{\textbf{28.57}} & \rev{\textbf{63.64}} & \rev{\textbf{84.21}} & \rev{-} & \rev{\textbf{94.37}} \\
    \midrule
    \rev{\multirow{7}{*}{\shortstack[l]{Claude-Sonnet\\4.6}}} & \rev{Zero-Shot} & \rev{45.33} & \rev{83.76} & \rev{32.56} & \rev{34.49} & \rev{29.48} & \rev{58.14} & \rev{-} & \rev{55.00} & \rev{25.64} & \rev{4.96} & \rev{-} & \rev{62.64} \\
     & \rev{Few-Shot} & \rev{60.33} & \rev{88.81} & \rev{36.56} & \rev{37.91} & \rev{34.33} & \rev{67.61} & \rev{-} & \rev{60.00} & \rev{24.24} & \rev{10.81} & \rev{-} & \rev{77.66} \\
     & \rev{Chain-of-Thought} & \rev{67.33} & \rev{90.33} & \rev{36.48} & \rev{35.66} & \rev{31.63} & \rev{71.88} & \rev{-} & \rev{51.43} & \rev{0.00} & \rev{15.58} & \rev{-} & \rev{82.49} \\
     & \rev{Mini-SWE-Agent} & \rev{80.67} & \rev{92.29} & \rev{47.60} & \rev{51.28} & \rev{48.03} & \rev{70.97} & \rev{-} & \rev{79.07} & \rev{40.00} & \rev{53.85} & \rev{-} & \rev{92.31} \\
     & \rev{OpenHands} & \rev{70.67} & \rev{91.67} & \rev{39.93} & \rev{42.80} & \rev{37.68} & \rev{63.64} & \rev{-} & \rev{51.43} & \rev{33.33} & \rev{32.43} & \rev{-} & \rev{82.90} \\
     & \rev{Claude Code} & \rev{68.67} & \rev{90.95} & \rev{37.62} & \rev{35.30} & \rev{33.52} & \rev{61.54} & \rev{-} & \rev{47.06} & \rev{28.57} & \rev{16.67} & \rev{-} & \rev{80.82} \\
     & \rev{\textbf{\tool}} & \rev{\textbf{88.33}} & \rev{\textbf{97.05}} & \rev{\textbf{62.40}} & \rev{\textbf{72.33}} & \rev{\textbf{65.95}} & \rev{\textbf{84.85}} & \rev{-} & \rev{\textbf{87.50}} & \rev{\textbf{63.16}} & \rev{\textbf{81.82}} & \rev{-} & \rev{\textbf{94.33}} \\
    \midrule
    \rev{\multirow{6}{*}{\shortstack[l]{Gemini-3.1\\Pro}}} & \rev{Zero-Shot} & \rev{55.33} & \rev{80.10} & \rev{45.56} & \rev{52.64} & \rev{42.64} & \rev{68.85} & \rev{-} & \rev{63.16} & \rev{32.26} & \rev{11.36} & \rev{-} & \rev{72.82} \\
     & \rev{Few-Shot} & \rev{71.00} & \rev{84.95} & \rev{44.43} & \rev{40.22} & \rev{39.43} & \rev{66.67} & \rev{-} & \rev{57.14} & \rev{33.33} & \rev{33.33} & \rev{-} & \rev{85.51} \\
     & \rev{Chain-of-Thought} & \rev{78.00} & \rev{89.10} & \rev{46.66} & \rev{41.62} & \rev{42.26} & \rev{77.19} & \rev{-} & \rev{55.56} & \rev{45.71} & \rev{27.27} & \rev{-} & \rev{90.09} \\
     & \rev{Mini-SWE-Agent} & \rev{57.00} & \rev{64.14} & \rev{59.67} & \rev{30.02} & \rev{38.95} & \rev{59.57} & \rev{-} & \rev{52.94} & \rev{34.78} & \rev{50.00} & \rev{-} & \rev{75.38} \\
     & \rev{OpenHands} & \rev{82.67} & \rev{95.14} & \rev{53.45} & \rev{41.39} & \rev{42.23} & \rev{78.12} & \rev{-} & \rev{28.57} & \rev{54.55} & \rev{42.86} & \rev{-} & \rev{91.49} \\
     & \rev{\textbf{\tool}} & \rev{\textbf{90.67}} & \rev{\textbf{95.19}} & \rev{\textbf{82.80}} & \rev{\textbf{64.68}} & \rev{\textbf{71.77}} & \rev{\textbf{82.76}} & \rev{-} & \rev{\textbf{81.82}} & \rev{\textbf{81.82}} & \rev{\textbf{94.12}} & \rev{-} & \rev{\textbf{95.24}} \\
    \midrule
    \rev{\multirow{6}{*}{\shortstack[l]{Qwen-3.5\\Plus}}} & \rev{Zero-Shot} & \rev{79.33} & \rev{94.00} & \rev{38.84} & \rev{26.25} & \rev{28.57} & \rev{58.62} & \rev{-} & \rev{34.48} & \rev{17.39} & \rev{0.00} & \rev{-} & \rev{89.53} \\
     & \rev{Few-Shot} & \rev{79.33} & \rev{94.10} & \rev{45.69} & \rev{29.26} & \rev{32.53} & \rev{64.29} & \rev{-} & \rev{28.57} & \rev{19.05} & \rev{26.67} & \rev{-} & \rev{89.16} \\
     & \rev{Chain-of-Thought} & \rev{78.67} & \rev{93.10} & \rev{43.63} & \rev{24.56} & \rev{26.74} & \rev{52.17} & \rev{-} & \rev{15.38} & \rev{30.77} & \rev{0.00} & \rev{-} & \rev{88.84} \\
     & \rev{Mini-SWE-Agent} & \rev{73.00} & \rev{77.67} & \rev{\textbf{56.29}} & \rev{22.61} & \rev{28.64} & \rev{37.21} & \rev{-} & \rev{40.00} & \rev{11.11} & \rev{20.00} & \rev{-} & \rev{75.97} \\
     & \rev{OpenHands} & \rev{79.67} & \rev{93.48} & \rev{40.26} & \rev{23.88} & \rev{26.69} & \rev{58.82} & \rev{-} & \rev{28.57} & \rev{10.53} & \rev{0.00} & \rev{-} & \rev{88.93} \\
     & \rev{\textbf{\tool}} & \rev{\textbf{83.33}} & \rev{\textbf{95.24}} & \rev{50.40} & \rev{\textbf{37.89}} & \rev{\textbf{42.38}} & \rev{\textbf{73.68}} & \rev{-} & \rev{\textbf{63.16}} & \rev{\textbf{31.58}} & \rev{\textbf{37.50}} & \rev{-} & \rev{\textbf{90.76}} \\
    \bottomrule
\end{tabular}%
}
\end{table*}

\rev{On SWE-Gym, \tool outperforms the strongest baseline by 9.02, 10.74, 15.26, and 17.76 EM points on Qwen, Claude, Gemini, and
  GPT, respectively. Although the ``Others'' category is especially challenging due to its rarity (0.7\%, 16 instances), \tool
  still achieves the best performance across all backbones.
On SWE-bench Multilingual, where the lower misalignment rate (24.3\%) leads to higher baseline EM, \tool still improves EM by
  3.66--8.00 points and achieves 42.38--71.85\% Macro F1, compared with 28.64--48.03\% for the strongest baseline. The
  metrics are less stable due to relatively few samples. For example, on Qwen, Mini-SWE-Agent attains
  higher Macro Precision (56.29\% vs.\ 50.40\%) through more conservative predictions, but this comes with lower Recall (22.61\%
  vs.\ 37.89\%), EM (73.00\% vs.\ 83.33\%), and Macro F1 (28.64\% vs.\ 42.38\%).}

\finding{\rev{\tool achieves the best binary Accuracy and multi-class EM across all four backbones on both datasets, leading the best baseline by 5.13--12.39 Accuracy and 9.02--17.76 EM points on SWE-Gym, and by 2.33--5.67 Accuracy and 3.66--8.00 EM points on SWE-bench Multilingual.}}

\FloatBarrier

     \subsection{RQ4: Ablation Study}
    \label{sec:rq4}

    In this RQ, we assess each component's contribution to \tool. We run ablations on the best-performing
  backbone Gemini-3.1-Pro-Preview, and use phase-wise analysis across all four backbones as complementary
  evidence.   
    \subsubsection{Approach}
    \label{sec:rq4_approach}

    \rev{To assess component contributions, we construct ten ablated variants on SWE-Gym using Gemini-3.1-Pro-Preview. The variants cover three granularities. First,
  phase-level ablations remove all Phase~I subagents, the Phase~II coordinator, or the
  Phase~III code validator. Second, single-subagent ablations remove the Issue Analyzer, PR
  Scope Analyzer, or PR Connection Analyzer while keeping the other two Phase~I subagents.
  Third, multi-component ablations remove pairs of Phase~I subagents or remove both the
  coordinator and code validator. We compare each variant with the full \tool pipeline using
  Binary Accuracy and Exact Match.}
    
    \subsubsection{Ablation Results}
    \label{sec:rq4_results}

\begin{table}[!t]
\centering
\caption{Ablation study results (backbone: Gemini-3.1-Pro-Preview). BA means Binary Accuracy; EM means Exact Match. $\Delta$ denotes the change relative to the full pipeline.}
\label{tab:rq4_ablation}
\setlength{\tabcolsep}{5pt}
\small
\resizebox{0.96\columnwidth}{!}{%
\begin{tabular}{p{4cm} cccc}
    \toprule
    \textbf{Variant} & \textbf{BA} & \textbf{$\Delta$} & \textbf{EM} & \textbf{$\Delta$} \\
    \midrule
    \textbf{\tool (full)} & \textbf{92.12} & --- & \textbf{84.66} & --- \\
    \midrule
    \rev{w/o Phase I (all 3 subagents)} & \rev{67.83} & \rev{$-$24.29} & \rev{64.78} & \rev{$-$19.88} \\
    w/o Phase II (Coordinator) & 87.41 & $-$4.71 & 77.44 & $-$7.22 \\
    w/o Phase III (Code Validator) & 88.97 & $-$3.15 & 79.00 & $-$5.66 \\
    \midrule
    w/o Issue Analyzer & 84.58 & $-$7.54 & 72.64 & $-$12.02 \\
    w/o PR Scope Analyzer & 73.26 & $-$18.86 & 61.16 & $-$23.50 \\
    w/o PR Conn. Analyzer & 85.77 & $-$6.35 & 74.82 & $-$9.84 \\
    \rev{w/o Issue \& PR Scope Analyzers} & \rev{59.32} & \rev{$-$32.80} & \rev{53.52} & \rev{$-$31.14} \\
    \rev{w/o Issue \& PR Conn. Analyzers} & \rev{81.65} & \rev{$-$10.47} & \rev{71.25} & \rev{$-$13.41} \\
    \rev{w/o PR Scope \& PR Conn. Analyzers} & \rev{67.91} & \rev{$-$24.21} & \rev{59.40} & \rev{$-$25.26} \\
    \rev{w/o Coordinator \& Code Validator} & \rev{85.60} & \rev{$-$6.52} & \rev{77.21} & \rev{$-$7.45} \\
    \bottomrule
\end{tabular}%
}
\end{table}

    Table~\ref{tab:rq4_ablation} reports the ablation results for the ten variants on Gemini-3.1-Pro-Preview.

    \rev{Removing all Phase~I subagents causes the largest drop ($-$24.29 BA, $-$19.88 EM), confirming the subagent ensemble as the detection backbone.} Removing the coordinator ($-$4.71 BA, $-$7.22 EM) or the code validator ($-$3.15 BA, $-$5.66 EM) causes smaller but meaningful degradation. EM drops exceed BA drops by 1.5$\times$ and 1.8$\times$ respectively, indicating that Phases~II and~III primarily refine label-level precision.

    Among individual agents, the PR Scope Analyzer has the greatest impact ($-$18.86 BA, $-$23.50 EM), consistent with the high prevalence of SC+UL (33.5\% of misaligned instances). The Issue Analyzer and PR Connection Analyzer have smaller but still substantial effects ($-$12.02 and $-$9.84 EM), matching the lower prevalence of IS (14.0\%) and DP+FP (10.7\%). \rev{Pairwise removals show complementarity, since combined drops exceed either corresponding individual drop.}

    \subsubsection{Phase-Wise Progression Analysis}
    \label{sec:rq4_phase}

\begin{figure}[!t]
\centering
\pgfplotsset{
    phase plot/.style={
        width=0.48\columnwidth, height=4cm,
        xlabel={Pipeline Phase},
        xtick={1,2,3},
        xticklabels={I,II,III},
        grid=major,
        grid style={dashed, gray!30},
        mark size=2pt,
        every axis plot/.append style={semithick},
        label style={font=\small},
        tick label style={font=\footnotesize},
        width=\linewidth,
    }
}
\ref{phase_legend}\\[2pt]
\begin{subfigure}[!t]{0.48\columnwidth}
\centering
\begin{tikzpicture}
\begin{axis}[
    phase plot,
    ylabel={Binary Accuracy (\%)},
    ymin=78, ymax=93,
    ytick={78,80,82,84,86,88,90,92},
]
\addplot[color=blue, mark=square*] coordinates {(1,87.1)(2,87.2)(3,88.1)};
\addplot[color=red, mark=triangle*] coordinates {(1,84.7)(2,85.4)(3,87.4)};
\addplot[color=olive, mark=diamond*] coordinates {(1,90.6)(2,91.5)(3,92.1)};
\addplot[color=purple, mark=o] coordinates {(1,80.1)(2,80.2)(3,80.5)};
\end{axis}
\end{tikzpicture}
\caption{Binary Accuracy}
\label{fig:ba_phases}
\end{subfigure}%
\hfill
\begin{subfigure}[!t]{0.48\columnwidth}
\centering
\begin{tikzpicture}
\begin{axis}[
    phase plot,
    ylabel={Exact Match (\%)},
    ymin=65, ymax=86,
    ytick={66,68,70,72,74,76,78,80,82,84,86},
    legend to name=phase_legend,
    legend columns=2,
    legend style={font=\scriptsize, draw=none, column sep=1pt, /tikz/every even column/.append style={column sep=1pt}, /tikz/every odd column/.append style={column sep=1pt}},
    width=\linewidth,
]
\addplot[color=blue, mark=square*] coordinates {(1,76.8)(2,77.2)(3,79.8)};
\addlegendentry{GPT-5.3-Codex}
\addplot[color=red, mark=triangle*] coordinates {(1,70.9)(2,72.1)(3,76.6)};
\addlegendentry{Claude-Sonnet-4.6}
\addplot[color=olive, mark=diamond*] coordinates {(1,81.2)(2,82.8)(3,84.7)};
\addlegendentry{Gemini-3.1-Pro}
\addplot[color=purple, mark=o] coordinates {(1,69.2)(2,69.6)(3,70.6)};
\addlegendentry{Qwen-3.5-Plus}
\end{axis}
\end{tikzpicture}
\caption{Exact Match}
\label{fig:em_phases}
\end{subfigure}
\Description{Two line charts showing binary accuracy and exact match across PAIChecker's three phases for four language-model backbones. Both metrics increase monotonically from Phase I through Phase III.}
\caption{Binary Accuracy and Exact Match across pipeline phases. Both metrics improve monotonically.}
\label{fig:phase_plots}
\end{figure}
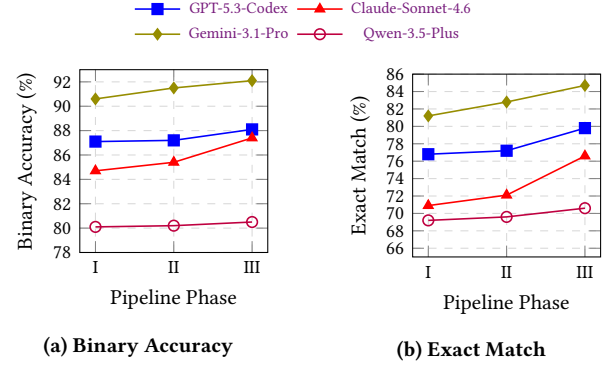

    Figure~\ref{fig:phase_plots} visualizes how BA and EM evolve across the three
  pipeline phases for all four backbones. Both metrics improve monotonically
  across all phases and backbones. Phase~I$\to$II gains are moderate and
  backbone-dependent (up to +1.6 EM for Gemini), reflecting the coordinator's
  conservative reconciliation role. Phase~II$\to$III gains are consistently
  larger, with EM improvements disproportionately exceeding BA gains across all
  backbones.

\finding{\rev{Phase~I provides the main detection signal, while Phases~II and~III primarily refine label-level accuracy. Individual and pairwise removals show that the three Phase~I subagents contribute complementary evidence rather than substitutable signals.}}

    \subsection{RQ5: Label Changes Analysis}
    \label{sec:rq5}

    This RQ examines how and why labels change between pipeline phases, analyzing what types of corrections each phase contributes and whether the multi-phase design yields net positive refinement.

    \subsubsection{Approach}
    \label{sec:rq5_approach}

    We track label changes between phases and classify each change as \textit{IC} (Incorrect$\to$Correct) or \textit{CI} (Correct$\to\allowbreak $Incorre-\allowbreak ct). We measure these transitions for Phase~I$\to$II and Phase~II$\allowbreak \to$III across all backbones, then manually inspect representative cases to explain the correction mechanisms.

    \subsubsection{Results}
    \label{sec:rq5_results}

\begin{table}[!t]
\centering
\caption{Label change statistics per phase per backbone. IC = corrections (Incorrect$\to$Correct), CI = regressions (Correct$\to$Incorrect). Percentages are relative to total label changes at each phase.}
\label{tab:self_correction}
\setlength{\tabcolsep}{3pt}
\small
\resizebox{0.96\columnwidth}{!}{%
\begin{tabular}{l cc c cc c}
    \toprule
    & \multicolumn{3}{c}{\textbf{Phase I$\to$II (Coordinator)}} & \multicolumn{3}{c}{\textbf{Phase II$\to$III (Code Validator)}} \\
    \cmidrule(lr){2-4} \cmidrule(lr){5-7}
    \textbf{Backbone} & \textbf{IC} & \textbf{CI} & \textbf{Total} & \textbf{IC} & \textbf{CI} & \textbf{Total} \\
    \midrule
    GPT-5.3-Codex & 11 (91.7\%) & 1 (8.3\%) & 12 & 100 (72.5\%) & 38 (27.5\%) & 138 \\
    Claude-Sonnet-4.6 & 74 (63.2\%) & 43 (36.8\%) & 117 & 125 (88.7\%) & 16 (11.3\%) & 141 \\
    Gemini-3.1-Pro & 58 (75.3\%) & 19 (24.7\%) & 77 & 67 (76.1\%) & 21 (23.9\%) & 88 \\
    Qwen-3.5-Plus & 21 (63.6\%) & 12 (36.4\%) & 33 & 40 (72.7\%) & 15 (27.3\%) & 55 \\
    \bottomrule
\end{tabular}%
}
\end{table}

    As shown in Table~\ref{tab:self_correction}, IC consistently outnumbers CI at both phases across all backbones. Phase~I$\to$II changes vary more in volume and IC rate, whereas Phase~II$\to$III changes are larger and remain mostly corrective. 

 \textbf{Phase~I$\to$II.}\quad Label changes mainly take two forms: adding \textit{Others} and dropping labels whose cited evidence does not support the assigned pattern. The first often occurs in three scenarios: (1)~\textit{partial implementation with altered behavior}, where the PR substitutes a requested behavior; (2)~\textit{partial implementation with omitted behavior}, where only part of the request is implemented; and (3)~\textit{completely mismatched pairs}, where the PR and issue are unrelated. The second reflects the coordinator's text-level self-correction: aggregating subagent evidence helps identify reasoning--label inconsistencies. For example, Figure~\ref{fig:motivation} shows a case where the reasoning indicates an unrelated PR, contradicting SC. Similarly, in \texttt{pandas-dev\_\_pandas-54451}, Phase~I flagged SC because the PR references both issue \#54617 and \#54167. The coordinator cross-referenced the PR Connection Analyzer, found the multi-issue reference to be a typo rather than SC, and correctly dropped SC.

\textbf{Phase~II$\to$III.}\quad Label changes at this phase only drop labels whose textual claims lack implementation support. For instance, in \texttt{python\_\_mypy-11632}, Phases~I and~II flagged SC because the PR claims to fix four issues. The code validator found from the diff that these issues are duplicates with one shared root cause, and correctly dropped SC, producing an exact match.

    \finding{IC consistently outnumbers CI across both phases and all backbones, confirming net positive label refinement. Phase~I$\to$II mainly performs text-level evidence re-verification, while Phase~II$\to$III performs implementation-level verification.}

    \section{Discussion}
    \label{sec:discussion}

\noindent\textbf{Real-World Relevance.}\quad
\rev{To test \tool on live GitHub PR--Issue pairs, we sampled 200 linked open PRs posted after January 2026 from nine popular, multilingual repositories. \tool flagged 78 potential misalignments, which we reported as GitHub comments for maintainer confirmation or dispute. As of writing, developers have responded to 17 reports and confirmed 16 (94\% of responses), suggesting that the detections transfer from curated benchmarks to development activity. }

\noindent\textbf{Token Consumption and Cost.}\quad
Across backbones, \tool averages 42K--59K tokens per instance. Processing SWE-Gym costs \$0.08--\$0.23 per instance. \rev{The code validator consumes about 40\% of tokens and usually adds \$0.02--\$0.10, although large patches can push total per-instance cost above \$1. Each instance completes in 30--100\,s, and independent instances can run in parallel.}

\noindent\textbf{Backbone Mixing.}\quad
Our evaluation uses one backbone for all agents. Assigning different models to different agents (\eg a reasoning-strong coordinator and code-proficient validator) may improve accuracy or reduce cost, but introduces a combinatorial search space.

\noindent\textbf{PR Description--Code Misalignment.}\quad
\tool filters PR claims unsupported by code, but code may silently introduce unrelated changes despite an aligned description. Because \tool also consumes review discussion, such unmentioned changes should be rare with negligible impact on the overall results.

    \section{Threats to Validity}
    \label{sec:threats}

    \textbf{Tool-Dependent Task Accessibility.}\quad
    \rev{Task validity may depend on the tools available to an agent. For example, an issue that relies on an external link is valid under our definition if developers do not raise questions. However, it becomes invalid for agents without web search tools. Our analysis does not evaluate task validity under all possible tool configurations, which is a limitation of this study.}

    \textbf{Internal Validity.}\quad
    The primary threat is annotation subjectivity.
We mitigate this through (1) a detailed annotation guideline with explicit boundary criteria, (2) dual-annotator independent labeling,
and consensus resolution with a third annotator for disagreements.

    \textbf{External Validity.}\quad
    Both datasets draw from a limited set of
GitHub repositories. However, the misalignment patterns stem
from general process-level factors such as how rigorously teams enforce issue--PR traceability, rather than language or domain-specific
features, suggesting the taxonomy transfers to repositories with
similar development workflows. 

    \section{Related Work}
    \label{sec:related}

    \textbf{SWE-Bench-Like Benchmarks.}\quad
    SWE-bench~\cite{swebench} introduced real-world GitHub issue resolution as an LLM evaluation paradigm. Its derivatives add human validation~\cite{swebenchverified}, live or automated extraction~\cite{swebenchlive,swerebench}, synthetic instances~\cite{swesmith}, runnable training data~\cite{swegym}, harder tasks~\cite{deng2025swepro}, and further language, performance, evolution, workflow, and scale dimensions~\cite{swepoly,he2025sweperf,wang_swe-bench_2025,thai_swe-evo_2025,xie_swe-fixer_2025,zhang_swe-flow_nodate,chen_swe-universe_2026,tao_swe-lego_2026}.

    \textbf{Benchmark Quality and Reliability.}\quad
    Benchmark reliability is a longstanding concern, with prior work highlighting test oracle insufficiency~\cite{utboost,wang_are_2025}, dataset contamination~\cite{swebenchverifiednolonger}, and solution leakage~\cite{aleithan_swe-bench_2024}. SWE-ABS~\cite{sweabs} strengthens test suites adversarially, while SPICE~\cite{oliva_spice_2025} automates labeling for SWE-bench-style datasets.

    \rev{\textbf{Comparison with OpenAI's Analysis.}\quad OpenAI's SWE-bench analysis~\cite{swebenchverifiednolonger} focuses on contamination and test coverage. This overlaps with parts of our taxonomy, such as PR Scope Creep, but not the construction-level root cause: a linked PR may not be intended to fix only the linked issue. Our taxonomy also covers cases beyond test coverage, such as defective PRs used as gold patches.}

    \rev{\textbf{Comparison With SPICE.}\quad SPICE~\cite{oliva_spice_2025} transfers SWE-bench Verified quality criteria to SWE-Gym, whereas \tool audits PR--Issue alignment. Among 644 shared instances, SPICE flags 143 and \tool flags 281, with 78 in common. \texttt{dask\_\_dask-7305} is \tool-only because its gold PR introduces a regression (DP); \texttt{dask\_\_dask-6862} is SPICE-only because an API-name difference preserves the requested semantics; and both flag \texttt{conan-io\_\_c-\allowbreak onan-16028}, whose PR adds unrelated utilities.}
    
    \section{Conclusion}
    \label{sec:conclusion}

This paper identifies PR-Issue misalignment as a systematic con-\allowbreak struction-level problem in SWE-bench-like benchmarks that directly
undermines evaluation fairness and reliability. Manual analysis
of all 500 SWE-bench Verified instances reveals 13.6\% misalignment across five patterns and 11 fine-grained scenarios. To enable scalable detection, we propose \tool, a three-phase multi-agent checker
  combining pattern-specific detection, beyond-taxonomy synthesis, and code validation. \rev{Evaluated on SWE-Gym and SWE-bench Multilingual, \tool reaches up to 92.12\% and 91.67\% accuracy, respectively.}

\begin{acks}
    This paper was supported by the Guangdong Basic and Applied Basic Research Foundation (No. 2024A1515010145) and the Shenzhen Science and Technology Program (Shenzhen Key Laboratory Grant No. ZDSYS20230626091302006).
\end{acks}

    \section*{Data Availability}
    \label{sec:data-availability}
    All data and artifacts are publicly available~\cite{replication}. 

    \bibliographystyle{ACM-Reference-Format}
    \bibliography{ref}
\end{document}